\documentclass[12pt]{article}
\usepackage{natbib,graphicx,setspace,amsmath,amssymb,subfigure,url,color}

\newcommand{\vc}{{\bf c}}

\newcommand{\vW}{{\bf W}}

\newcommand{\vE}{{\bf E}}
\newcommand{\vG}{{\bf G}}
\newcommand{\vF}{{\bf F}}

\newcommand{\vu}{{\bf u}}
\newcommand{\vv}{{\bf v}}
\newcommand{\vw}{{\bf w}}

\newcommand{\vA}{{\bf A}}
\newcommand{\vB}{{\bf B}}
\newcommand{\vC}{{\bf C}}
\newcommand{\vH}{{\bf H}}
\newcommand{\vI}{{\bf I}}

\newcommand{\vM}{{\bf M}}
\newcommand{\vN}{{\bf N}}
\newcommand{\vP}{{\bf P}}
\newcommand{\vQ}{{\bf Q}}

\newcommand{\vU}{{\bf U}}
\newcommand{\vV}{{\bf V}}
\newcommand{\vX}{{\bf X}}
\newcommand{\vY}{{\bf Y}}
\newcommand{\vZ}{{\bf Z}}
\newcommand{\vD}{{\bf D}}

\newcommand{\vnull}{{\bf 0}}

\newcommand{\bay}{\begin{array}}
\newcommand{\eay}{\end{array}}

\newcommand{\bqa}{\begin{eqnarray*}}
\newcommand{\eqa}{\end{eqnarray*}}
\newcommand{\bqan}{\begin{eqnarray}}
\newcommand{\eqan}{\end{eqnarray}}
\newcommand{\bqt}{\begin{quote}}
\newcommand{\eqt}{\end{quote}}
\newcommand{\bt}{\begin{tabbing}}
\newcommand{\et}{\end{tabbing}}
\newcommand{\bit}{\begin{itemize}}
\newcommand{\eit}{\end{itemize}}
\newcommand{\ben}{\begin{enumerate}}
\newcommand{\een}{\end{enumerate}}
\newcommand{\beq}{\begin{equation}}
\newcommand{\eeq}{\end{equation}}
\newcommand{\bdefi}{\begin{definition}}
\newcommand{\edefi}{\end{definition}}
\newcommand{\bpro}{\begin{proposition}}
\newcommand{\epro}{\end{proposition}}
\newcommand{\bco}{\begin{corollary}}
\newcommand{\eco}{\end{corollary}}
\newcommand{\bdes}{\begin{description}}
\newcommand{\edes}{\end{description}}

\def\log{\hbox{log}}

\def\boxit#1{\vbox{\hrule\hbox{\vrule\kern6pt
          \vbox{\kern6pt#1\kern6pt}\kern6pt\vrule}\hrule}}

\def\bse{\begin{eqnarray*}}
\def\ese{\end{eqnarray*}}
\def\be{\begin{eqnarray}}
\def\ee{\end{eqnarray}}
\def\bq{\begin{equation}}
\def\eq{\end{equation}}

\newtheorem{proposition}{Proposition}

\newcommand{\blem}{\begin{lemma}}
\newcommand{\elem}{\end{lemma}}
\newcommand{\bthe}{\begin{theorem}}
\newcommand{\ethe}{\end{theorem}}

\newtheorem{definition}{Definition}[section]
\newtheorem{lemma}[definition]{Lemma}
\newtheorem{theorem}[definition]{Theorem}

\def\delete#1{\iffalse #1 \fi}

\def\bse{\begin{eqnarray*}}
\def\ese{\end{eqnarray*}}
\def\bee{\begin{enumerate}}
\def\eee{\end{enumerate}}
\def\bqe{\begin{eqnarray}}
\def\eqe{\end{eqnarray}}
\def\bed{\begin{description}}
\def\eed{\end{description}}
\def\bei{\begin{itemize}}
\def\eei{\end{itemize}}

\def\pmb#1{\setbox0=\hbox{#1}%
    \kern-.025em\copy0\kern-\wd0
    \kern.05em\copy0\kern-\wd0
    \kern-.025em\raise.0433em\box0 }
\def\pmbh#1#2{\setbox0=\hbox{#1}%
    \setbox1=\hbox{#2}%
    \kern-.025em\copy0\kern-\wd0
    \kern.05em\copy1\kern-\wd0
    \kern-.025em\raise.0433em\box0 }

\def\frac#1#2{{#1\over#2}}

\def\boxit#1{\vbox{\hrule\hbox{\vrule\kern6pt
   \vbox{\kern6pt#1\kern6pt}\kern6pt\vrule}\hrule}}

\def\listing#1{\vskip 4mm\begin{verbatim}\input#1 \vskip 4mm}
\def\thick#1{\hbox{\rlap{$#1$}\kern0.25pt\rlap{$#1$}\kern0.25pt$#1$}}

\def\pmbh{{\pmb h}}

\def\calA{{\cal A}}
\def\calB{{\cal B}}

\def\calF{{\cal F}}

\def\calH{{\cal H}}

\def\calM{{\cal M}}

\def\calO{{\cal O}}

\renewcommand\today{\ifcase\month\or
   Jan\or Feb\or Mar\or Apr\or May\or
   Jun\or Jul\or Aug\or Sep\or Oct\or Nov\or
   Dec\fi
   \space\number\day, \number\year}

\DeclareMathOperator*{\argmin}{arg\,min}
\def\boxit#1{\vbox{\hrule\hbox{\vrule\kern6pt
          \vbox{\kern6pt#1\kern6pt}\kern6pt\vrule}\hrule}}

\newtheorem{thm}{Theorem}
\newtheorem{lem}{Lemma}
\newtheorem{defi}{Definition}
\newtheorem{rmk}{Remark}
\newtheorem{prp}{Proposition}
\newtheorem{clm}{Claim}
\begin{document}

\title{Reduced-rank Regression in Sparse Multivariate Varying-Coefficient Models with High-dimensional Covariates \footnote{\baselineskip%
=10pt Heng Lian is Assistant Professor, Division of Mathematical Sciences, SPMS, Nanyang Technological University, Singapore (Email: henglian@ntu.edu.sg). Shujie Ma is Assistant Professor, Department of Statistics, University
of California-Riverside, Riverside, CA 92521 (Email: shujie.ma@ucr.edu).
Ma's research was partially supported
by NSF grant DMS 1306972.}}
\author{Heng Lian and Shujie Ma\\}
\date{}          
\maketitle
\begin{abstract}
In genetic studies, not only can the number of predictors obtained from microarray measurements be extremely large, there can also be multiple response variables. Motivated by such a situation, we consider semiparametric dimension reduction methods in sparse multivariate regression models. Previous studies on joint variable and rank selection  have focused on parametric models while here we consider the more challenging varying-coefficient models which make the investigation on nonlinear interactions of variables possible. Spline approximation, rank constraints and concave group penalties are utilized for model estimation. Asymptotic oracle properties of the estimators are presented. We also propose reduced-rank independent screening to deal with the situation when the dimension is so high that penalized estimation cannot be efficiently applied. In simulations, we show the advantages of simultaneously performing variable and rank selection. A real data set is analyzed to illustrate the good prediction performance when incorporating interactions between genetic variables and an index variable.

\vspace{0.5cm}


\vspace{0.5cm}

\noindent \textbf{Short title:} Reduced-rank regression in VC models
\end{abstract}

\newpage

\section{Introduction}

The Framingham Heart Study (FHS), started in 1940 and still continuing, is a project in health research to identify the common factors that contribute to cardiovascular diseases. We use SNP data on 6847 patients in the study. Moreover, there are $15$ phenotypes available on 325 patients, as shown in Table \ref{tab:response} in Section 4, which are used as response variables. After matching the SNP data with the phenotypes and deleting observations with missing values, there are 292 patients remaining in our study. Obviously, based on the descriptions on the phenotypes, they are naturally correlated. With a large number of SNPs (32164 SNPs for one particular chromosome that we focus on in our numerical illustrations), clearly it is crucial to identify a small number of them that are important in explaining the response variables. In particular, we are interested in how the genetic effect changes with the physical activity level of the patient for which a varying-coefficient structure is appropriate. Besides identifying important SNPs that interact with the index variable, it is also important to take into account the correlations of the responses in some way to construct a more parsimonious model.

In multivariate regression problems, we are given i.i.d. observations $(\vY_i, \vX_i), i=1,\ldots, n$ that are independently and identically distributed (i.i.d.), where $\vY_i=(Y_{i1},\ldots,Y_{iq})^T$ are $q$-dimensional responses and $\vX_i=(X_{i0}=1, X_{i1},\ldots,X_{ip})^T$ are $p+1$-dimensional predictors. The model posed is
\begin{equation}\label{eqn:linear}
\vY=\vX\vC+\vE,
\end{equation}
where $\vY=(\vY_1,\ldots,\vY_n)^T$, $\vX=(\vX_1,\ldots,\vX_n)^T$, $\vC$ is the $(p+1)\times q$ coefficient matrix to be estimated, and $\vE$ is the $n\times q$ noise matrix with independent rows. The ordinary least squares estimation method minimizes $\|\vY-\vX\vC\|^2$ which reduces to a separate linear regression for each response. Here we use $\|.\|$ to denote the Frobenius norm of a matrix, that is, $\| \vY-\vX\vC\| ^{2}={\rm tr}\{(\vY-\vX\vC)^T(\vY-\vX\vC)\}$.

With a large number of responses and/or predictors, which is the main focus of the present study, a more parsimonious model is needed to avoid unidentifiability, singularity, and overfitting. Two popular approaches for achieving parsimony in the context of multivariate regression are reduced-rank regression \citep{izenman75,bunea11,kunchen13}, which assumes that $\vC$ is of low rank, and sparse regression \citep{tibshirani96,fan01,zou06,zhaoyu06,zhanghuang08,bickel09,wang2011estimation}, which assumes that only a small subset of the collected predictors are relevant for prediction of the responses and this in multivariate regression is equivalent to saying that $\vC$ contains only a small number of nonzero rows. Combination of these two complementary constraints has also been proposed and analyzed in \cite{bunea12,kunchen12,chenhuang12} which shows improved performance than using only one of the constraints in high-dimensional situations.

However, as have been demonstrated in many papers for sparse univariate regression including \cite{wang08,wang09,huang10}, linear models are sometimes not flexible enough to achieve satisfactory prediction performance in real applications due to their stringent parametric assumptions. To avoid fully nonparametric regression with limited data which resides at the other extremum of regression modelling, many nonparametrically structured models were proposed in the literature of sparse regression that can model flexibly nonlinear covariate effects or covariate interactions in a parsimonious way \citep{huang10,wang08,wang09}.
As mentioned above, in this study, we focus on varying-coefficient modelling for multivariate regression which is given by
$$E[Y_{il}|\vX_i]=f^{(l)}_0(T_i)+\sum_{j=1}^pf_j^{(l)}(T_{i})X_{ij},$$
for nonparametric coefficients $f_j^{(l)}$, $0\le j\le p$, $1\le l\le q$ and index variable $T$.

We next propose an estimation procedure that simultaneously removes insignificant predictors and uses rank constraint to capitalize on similarities among different responses. Our approach is based on polynomial spline approximation for the nonparametric coefficient functions, with concave penalties to shrink blocks of the spline coefficients to zero. Then we use explicit multiplicative decomposition of the coefficient matrix to take into account the rank constraint, similar to \cite{bunea12,chenhuang12}. Compared to those previous parametric models, dealing with the varying coefficient model is more challenging both theoretically and computationally, since estimation in each nonparametric function involves a diverging number of nuisance parameters and moreover the number of coefficient functions increases with the dimensions of both the predictors and responses. 
For specificity, we use the smoothly clipped absolute deviation (SCAD) penalty and note that other concave penalties will produce similar empirical results in our experiences. We show the nonparametric oracle property of the estimator, that is, the irrelevant variables are consistently removed and nonzero functions are estimated with the same rate as when only the relevant variables are included in the model. 

In Section 2.2, we propose an algorithm to estimate the coefficient matrix. However, due to algorithmic limitations, penalized reduced-rank regression can only handle hundreds of predictors in our implementation. Hence, it cannot be directly applied to the FHS data set which contains tens of thousands of SNPs. In Section 3, we propose a semiparametric reduced-rank screening procedure that uses conditional marginal correlations to reduce the number of SNPs before applying penalized regression.
Unlike previous screening studies that rely on closed-form expression of the estimator to derive the sure screening property, investigations on reduced-rank estimator which has no closed-form expression pose some theoretical difficulties. Section 4 is devoted to numerical studies. Our simulations demonstrate that taking into account both rank constraint and predictor sparsity jointly is significantly better than using only one of them. We also use simulations to investigate the proposed independence screening procedure and show that reduced-rank screening may have some advantages over full-rank screening although the improvements are relatively small. Our analysis on the FHS data set further confirms that the varying-coefficient model is better in prediction accuracy than linear sparse reduced-rank models. We conclude with some discussions in Section 5. The technical proofs for the main results are deferred to the Appendix.

\section{Penalized Estimation with Polynomial Splines}

\subsection{Setup and Estimation Approach}
Without loss of generality, we assume that the distribution of $T_{i}$ is supported on $[0,1]$. We use polynomial splines to approximate the components. Let $\tau_0=0<\tau_1<\cdots<\tau_{K'}<1=\tau_{K'+1}$ be a partition of $[0,1]$ into subintervals $[\tau_k,\tau_{k+1}),k=0,\ldots,K'$ with $K'$ internal knots. We only restrict our attention to equally spaced knots although data-driven choice can be considered such as putting knots at certain sample quantiles of the observed covariate values. A polynomial spline of order $m$ is a function whose restriction to each subinterval is a polynomial of degree $m-1$ and globally $m-2$ times continuously differentiable on $[0,1]$. The collection of splines with a fixed sequence of knots has a normalized B-spline basis $\{B_{1}(x),\ldots,B_{K}(x)\}$ with $K=K'+m$. Using spline expansions, we can approximate the components by $f_j^{(l)}(x)\approx \sum_k c^{(l)}_{jk}B_{k}(x)$. Note that it is possible to specify different $K$ for each response or even for each coefficient but we assume they are the same for simplicity.

The conditional expectation of the observed responses $\vY$ is the $n\times q$ matrix
\begin{equation*}
\vM=\left(\begin{array}{ccc}
       \sum_{j=0}^p f^{(1)}_j(T_1)X_{1j} & \ldots & \sum_{j=0}^p f^{(q)}_j(T_1)X_{1j}\\
        \vdots &\vdots & \vdots\\
       \sum_{j=0}^p f^{(1)}_j(T_n)X_{nj} & \ldots & \sum_{j=0}^p f^{(q)}_j(T_n)X_{nj}
          \end{array}\right).
\end{equation*}
We can write $\vM=\vM_0+\cdots +\vM_p$ with
\begin{equation*}
\vM_j=\left(\begin{array}{ccc}
       f^{(1)}_j(T_1)X_{1j} & \ldots & f^{(q)}_j(T_1)X_{1j}\\
        \vdots &\vdots & \vdots\\
       f^{(1)}_j(T_n)X_{nj} & \ldots & f^{(q)}_j(T_n)X_{nj}
          \end{array}\right).
\end{equation*}
In linear multivariate regression (\ref{eqn:linear}), the rank constraint takes the form $rank(\vC)\le r$ for $r \le \min\{p+1,q\}$. Certainly $rank(\vC)\le r$ implies $rank(\vX\vC)\le r$ and the converse is not necessarily true. However, the following proposition shows that the constraint $rank(\vC)\le r$ is actually equivalent to $rank(\vX\vC)\le r$ in estimation.

\begin{prp}  If $rank(\vX\vC)\le r$, then there is a matrix $\vC'$ with $rank(\vC')\le r$ and $\vX\vC=\vX\vC'$.
\end{prp}
\textbf{Proof.} Suppose $rank(\vX\vC)=r'\le r$, we can write $\vX\vC=\vB\vA^T$ where $\vB\in R^{(p+1)\times r'}$ and $\vA\in R^{q\times r'}$ with $rank(\vB)=rank(\vA)=r'$. Since the dimensions of $span\{\vB\vA^T\}$ (the space spanned by the columns of the matrix) and $span\{\vB\}$ are both $r'$, we have $span\{\vB\}=span\{\vB\vA^T\}=span\{\vX\vC\}\subseteq span\{\vX\}$ and thus there is a matrix $\vD$ such that $\vB=\vX\vD$. This implies $\vX\vC=\vX\vD\vA^T$ and the matrix $\vC':=\vD\vA^T$ has rank at most $r'$ since $rank(\vA)=r'$. \hfill $\Box$


For the multivariate varying-coefficient regression, one can naturally put the constraint $rank(\vM)\le r $. This means that there are $r $ columns of $\vM$ (corresponding to $r $ response variables) such that each of the $q$ columns of $\vM$  is actually a linear combination of those $r $ columns (effectively there are only $r $ independent responses). 
We further define
\[\vZ_j=\left(\begin{array}{cccc}
	B_{1}(T_1)X_{1j}&B_{2}(T_1)X_{1j}&\cdots&B_{K}(T_1)X_{1j}\\
        \vdots&\vdots&\vdots&\vdots\\
	B_{1}(T_n)X_{nj}&B_{2}(T_n)X_{nj}&\cdots&B_{K}(T_n)X_{nj}\\
	\end{array}\right)_{n\times K},\]
$\vZ=(\vZ_0,\ldots,\vZ_p)$. We collect the spline coefficients into a $(p+1)K\times q$ matrix $\vC$ with
$\vC=(\vC_0^T,\ldots,\vC_p^T)^T$ and
\begin{equation*}
\vC_j=\left(\begin{array}{ccc}
       c^{(1)}_{j1} &\ldots &c^{(q)}_{j1}\\
        \vdots &\vdots & \vdots\\
       c^{(1)}_{jK} &\ldots &c^{(q)}_{jK}\\
      \end{array}\right)
\end{equation*}
is the $K\times q$ matrix associated with predictor $j$. The rank-constrained minimization problem is
\begin{equation*}
\hat\vC=\argmin_{rank(\vZ\vC)\le r} \|\vY-\vZ\vC\|^2,
\end{equation*}
with rank constraint on $\vZ\vC$. It looks more convenient, especially considering the sparsity penalty on $\vC_j$ to be imposed below, to use rank constraint on $\vC$, as we will adopt in this paper. Thus the minimization above is performed with $rank(\vC)\le r$ for some integer $r$.

Obviously when $p$ is large, the rank constraint alone is not sufficient to obtain a parsimonious model. This can be seen from the fact that when $q=1$, the univariate regression model has at most rank one but still suffers from high dimensionality, if estimated without other constraints or penalties.
Thus we are interested in an even more parsimonious model where many predictors are irrelevant for prediction by some criteria. Without loss of generality, we assume only the first $s$ predictors are useful for prediction. Mathematically, we assume that $f_j^{(l)}=0$ for all $l$ when $j>s$, and $f_j^{(l)}\neq 0$ for at least some $l$ when $j\le s$. Such information on the ordering of predictors is only assumed in theoretical derivations later for notational convenience and not used for estimation.

Our finally proposed estimation procedure for joint variable selection and reduced-rank regression is
\begin{equation}\label{eqn:min}
\hat{\vC}=\argmin_{\vC: rank(\vC)\le r}\|\vY-\vZ\vC\|^2+n\sum_{j=1}^pp_\lambda(\|\vC_j\|),
\end{equation}
where $\lambda$ (as well as $r$) is a tuning parameter. There is more than one way to specify the penalty functions and here we only focus on the SCAD penalty function \citep{fan01}, defined by its first derivative
\[
\dot{p}_{\lambda}(x)=\lambda\left\{I(x\le\lambda)+\frac{(a\lambda-x)_+}{(a-1)\lambda}I(x>\lambda)\right\},
\]
with $a>2$ and $p_\lambda(0)=0$. We will use $a=3.7$ as suggested in \cite{fan01}. Other choices of penalty, such as adaptive lasso \citep{zou06} or minimax concave penalty \citep{zhang2010nearly}, are expected to produce similar results in both theory and practice.
Note that unlike most other penalized estimation in the statistics literature, here we are penalizing the Frobenius norm of a matrix, $\|\vC_j\|$, which is associated with the $j$-th predictor. This does not make an essential difference though, since the matrix can be vectorized as we mention in the following subsection.

\subsection{Computational Algorithm}
The computational algorithm we use for solving (\ref{eqn:min}) is similar to that used in \cite{bunea12,chenhuang12} with some modifications necessary as explained below.

First, to take into account the constraint $rank(\vC)\le r$, we parametrize $\vC=\vB\vA^T$ with $\vB$ a $(p+1)K\times r$ matrix and $\vA$ a $q\times r$ matrix satisfying $\vA^T\vA=\vI$. The orthonormality constraint of $\vA$ does not uniquely identify the parameters since $\vB\vA^T=(\vB\vQ)(\vQ^T\vA^T)$ for any orthogonal $r\times r$ matrix $\vQ$, but this does not prevent one from developing an algorithm using this parameterization that converges to a stationary point of (\ref{eqn:min}).

With the parameterization $\vC=\vB\vA^T$, we use the following alternate search algorithm to find the solution.
\begin{enumerate}
\item Let $\vC^{ols}=(\vZ^T\vZ)^{-}\vZ^T\vY$, where $()^{-}$ denotes the pseudoinverse. Perform singular value decomposition $\vC=\vU\vD\vV^T$. Initialize $\vB$ as the first $r$ columns of $\vU\vD$ and $\vA$ as the first $r$ columns of $\vV$.
\item For fixed $\vB$, we obtain new value of $\vA$ by solving $\min_{\vA: \vA^T\vA=\vI}\|{\vY}-\vZ\vB\vA^T\|^2$. This minimization problem has a closed-form solution.  In fact, if we have the SVD $\vY^T\vZ\vB=\vU\vD\vV^T$, we set $\vA=\vU\vV^T$.
\item For fixed $\vA$, we obtain new value of $\vB$ by solving $\min_{\vB}\|{\vY}-\vZ\vB\vA^T\|^2+n\sum_jp_\lambda(\|\vB_j\|)$, where $\vB=(\vB_0^T,\ldots,\vB_p^T)^T$ with $\vB_j$ being $K\times r$ matrices (we note that $\|\vB_j\|=\|\vC_j\|$ due to orthonormality of $\vA$). Noting that the minimization problem is nonconvex, for the purpose of discussion of convergence that follows, we assume that a local minimizer of $\vB$ can be obtained in this step.
\item If some convergence criterion is met, STOP. Otherwise, go back to step 2.
\end{enumerate}
In Step 3 above, one can use local quadratic approximation (an instantiation of MM algorithm as pointed out by \cite{zou08}) and iteratively approximate the penalty by
$$p_{\lambda}(\|\vB_j\|)\approx p_{\lambda}(\|\vB_j^{(0)}\|)+\frac{1}{2}\frac{\dot{p}_{\lambda}(\|\vB_j^{(0)}\|)}{\|\vB_j^{(0)}\|}{\{\|\vB_j\|^2-\|\vB_j^{(0)}\|^2\}},$$
where $\vB_j^{(0)}$ represents the current estimate of $\vB_j$. Due to the focus on large $p$ in this study, iteratively solving for this quadratic function of $\vB_j$ for all $j$ simultaneously is not efficient. We thus adopt the strategy of solving
\begin{equation}\label{eqn:minj}
\min_{\vB_j}\|{\vY}\vA-\sum_{j':j'\neq j}\vZ_{j'}\vB_{j'}-\vZ_j\vB_j\|^2+np_\lambda(\|\vB_j\|),
\end{equation}
(again using local quadratic approximation for this specific $j$), for $j$ from $0$ to $p$ in turn, in an iterative way (note when $j=0$ there is no penalty). Even though the objective function above is represented in terms of matrix $\vB_j$, during implementation $\vB_j$  can be vectorized using the fact that ${\rm vec}(\vZ_j\vB_j)=(\vI\otimes \vZ_j){\rm vec}(\vB_j)$ and thus (\ref{eqn:minj}) is equivalent to
\begin{equation*}
\min_{{\rm vec}(\vB_j)}\|{\rm vec}({\vY}\vA-\sum_{j':j'\neq j}\vZ_{j'}\vB_{j'})-(\vI\otimes\vZ_j){\rm vec}(\vB_j)\|^2+np_\lambda(\|{\rm vec}(\vB_j)\|),
\end{equation*}
which takes the same form as the group SCAD penalty used for example in \cite{wang08}. Computationally this is more complicated compared to the computation of linear model in \cite{chenhuang12}, for which (\ref{eqn:minj}) has a closed form solution when adaptive group lasso penalty is used for each $j$. For nonparametric problem, even if we adopt the adaptive group lasso penalty (instead of the concave SCAD penalty), we do not have a close-form solution and thus local approximation is necessary in our case. Although we used LQA algorithm here, other algorithm such as locally linear approximation (LLA) can be considered \citep{zou08}. However, unlike the linear problem studied in \cite{zou08}, even after using LLA, the lasso-type problem with a group penalty is still not trivial to solve. Thus we only used LQA algorithm in this paper.

We now discuss briefly the convergence property of the algorithm. We say a function $f(\vA,\vB)$ is biconvex if it is convex for either one of the arguments when the other is fixed. Although our objective function (\ref{eqn:min}) is not biconvex when we write $\vC$ as $\vB\vA^T$, we can borrow terminologies for biconvex functions in \citep{gorski07}.

\begin{defi}
We say $(\hat{\vA},\hat{\vB})$ is a local partial optimum of $f$ if $\hat{\vA}$ is a local minimizer of $f(\vA,\hat{\vB})$ and $\hat{\vB}$ is a local minimizer of $f(\hat{\vA},\vB)$.
\end{defi}
In general, under mild smoothness assumptions, a local minimizer is a local partial optimum and a local partial optimum is a stationary point. Following the same proof of Theorem 6 (i) in \cite{bunea12}, it can be shown that any accumulation point of the parameters obtained by the algorithm above is a local partial optimum. Although this convergence statement leaves many open questions such as when the local partial optimum is actually a local or even global optimum, and when the entire sequence will converge, general results seem hard to obtain to answer these theoretical questions. In our experience the algorithm always converges.

In practice, we need to choose some parameters including the spline order $m$, the number of basis $K$, the rank constraint bound $r$, and the regularization parameter $\lambda$. As commonly adopted, we fix $m=4$ (cubic splines) in all our numerical results.  To ease the computational burden, we fix $K = 5$ following \cite{huang10}. This
choice of $K$ is small enough to avoid overfitting in typical problems with sample size not too  small, and  big enough  to flexibly approximate many smooth functions. Finally, we use five-fold cross-validation to select the empirically more critical parameters $r$ and $\lambda$.

\subsection{Asymptotic Analysis}
In this subsection we study the asymptotic behavior of the estimator that allows $p$ and $q$ to grow with $n$. Our asymptotic results show that the estimator has the same convergence rate as when the zero rows of $\vB$ are known, and the zero rows can be consistently identified. Thus we can say the estimator has the nonparametric oracle property as defined in \cite{storlie2011surface}.

The following regularity conditions are used.
\begin{itemize}
\item[(c1)] The index variable $T$ has a continuous density supported on $[0,1]$ and the density is bounded away from zero and infinity on $[0,1]$. For theoretical simplicity, we also assume the predictors $X_j$ are bounded, although this can be relaxed at the cost of lengthier arguments.
\item[(c2)] The covariance matrix of $(X_j, 0\le j\le s)$ is bounded away from zero and infinity.
\item[(c3)] The noise matrix $\vE$ has i.i.d entries which has a subGaussian distribution.
\item[(c4)] The number of nonzero components $s$ is fixed. Let $f_{0j}^{(l)}, 0\le j\le p, 1\le l\le q$ be the true component functions. $f_{0j}^{(l)}\equiv 0$ for $j>s$.
\item[(c5)] For $g=f_{0j}^{(l)}, 0\le j\le s, 1\le l\le q$, $g$ satisfies a Lipschitz condition of order $d>1/2$: $|g^{(\lfloor d\rfloor)}(t)-g^{(\lfloor d\rfloor)}(s)|\le C|s-t|^{d-\lfloor d\rfloor}$, where $\lfloor d \rfloor$ is the biggest integer strictly smaller than $d$ and $g^{(\lfloor d\rfloor)}$ is the $\lfloor d\rfloor$-th derivative of $g$. The order of the B-spline used satisfies $m\ge d+2$.
\item[(c6)] $\|f_{0j}^{(l)}\|$, $1\le j\le s, 1\le l\le q$, are bounded away from zero.
\end{itemize}
These assumptions are common in the literature of sparse nonparametric models, see for example \cite{huang10}. In (c4), following \cite{huang10}, we assume the number of nonzero components is bounded and does not diverge with $n$. This is mainly due to technical reasons that we want to use the result that eigenvalues of  $\vZ_S^T\vZ_S$ are of order $n/K$ where $\vZ_S=(\vZ_1,\ldots,\vZ_s)^T$ (see Lemma A.1 in \cite{huang04}), which is reasonable only when $s$ is bounded. When $s$ diverges with $n$, it is more reasonable for example to assume the smallest eigenvalue of $\vZ_S^T\vZ_S$ has an order smaller than $n/K$ and the convergence rate will depend on this value. However, it seems hard to study the behavior of the eigenvalues when $s$ diverges.

Let $\vc_{0j}^{(l)}=(c_{0j1}^{(l)},\ldots,c_{0j}^{(q)})^T$ be the $K$-vector that satisfies $\|f_{0j}^{(l)}-\sum_{k=1}^K c_{0jk}^{(l)}B_{jk}\|_\infty=O(K^{-d})$ for $0\le j\le s, 1\le l\le q$ and $\vc_{0j}^{(l)}=\vnull$ for $j>s, 1\le l\le q$. Define $\vC_{0j}=(\vc_{0j}^{(1)},\ldots,\vc_{0j}^{(q)})$ and $\vC_0=(\vC_{00}^T,\ldots,\vC_{0p}^T)^T$.

\begin{thm}\label{thm:rates} (Convergence rates for estimation of $\vC$) Under conditions (c1)-(c6) and that $K\log  (pK)/n\rightarrow 0, K\rightarrow \infty$, $\lambda/\sqrt{qK}\rightarrow 0$, $\sqrt{nq\log(pK)}=o(n\lambda)$, $\sqrt{nq}(\sqrt{(q+K)r}+\sqrt{nq}K^{-d})=o(n\lambda)$, there is a local partial optimum of (\ref{eqn:min}) that satisfies
\begin{equation}\label{eqn:rates1}
\|\hat{\vC}-\vC_0\|^2=O_p\left(\frac{K}{n}\left((q+K)r+\frac{nq}{K^{2d}}\right)\right).
\end{equation}
The convergence rate of $\hat{\vC}$ implies the convergence of the coefficient functions
\begin{equation}\label{eqn:rates2}
\sum_{j=1}^p\sum_{l=1}^q \|\hat{f}_j^{(l)}-{f}_{0j}^{(l)}\|^2=O_p\left(\frac{(q+K)r}{n}+\frac{q}{K^{2d}}\right).
\end{equation}
\end{thm}


\begin{thm}\label{thm:vs} (Consistency of variable selection) Under the same assumptions as in Theorem \ref{thm:rates}, there is a local partial optimum of (\ref{eqn:min}) that satisfies $\vC_j\equiv \vnull$, $j>s$ (and achieving the convergence rate in Theorem \ref{thm:rates} at the same time).
\end{thm}
The proofs of the theorems are given in the Appendix. We note that even though $s$ is fixed, the rank $r$ can still diverge since it can be as large as $\min\{(s+1)K, q\}$. Also $K\log(pK)/n\rightarrow 0$ implies  $p\ll \exp
\{n/K\}$.

\section{Reduced-rank Independence Screening}
As we know, when the dimension of covariates is ultra-high, a screening procedure is necessary to screen out the completely irrelevant components. To take into account the fact that the multivariate response variables are related, we naturally consider a nonparametric reduced-rank independent screening procedure in the varying coefficient model, as an extension of the independent screening method with univariate responses proposed in \cite{fanmadai13} and \cite{liuliwu13}. Considering one covariate $X_j$ at a time, we fit the varying coefficient model (including an intercept)
$$\min_{rank((\vC_0^T,\vC_j^T))\le r_j} \|\vY-\vZ_0\vC_0-\vZ_j\vC_j\|^2.$$
For  simplicity of notation in this section we write $\bar\vZ_j=(\vZ_0,\vZ_j)$ and $\vH_j=(\vC_0^T,\vC_j^T)^T$. Denote the minimizer above by $\hat\vH_j$. Note that for different $j$ we can use a different $r_j$. Theoretically, we assume that $r_j$ is given in our asymptotic derivations below. In practice $r_j$ will be selected again by five-fold cross-validation. Let $\hat{\beta}_j=\|\bar\vZ_j\hat\vH_j\|^2/n$. We rank the covariates by decreasing values of $\hat{\beta}_j$. That is, the $j$-th covariate with larger $\hat{\beta}_j$ are considered to be more  important. More specifically, for some appropriately chosen threshold $\theta_n>0$, we will select the variables in the set $\hat{\calM}=\{j: \hat{\beta}_j\ge \theta_n\}$.

The population counterpart of $\hat{\beta}_j$ is $\beta_j=E\|\vu_j(T_i)+\vv_j(T_i)X_{ij}\|^2$, where $\vu_j=(u_j^{(1)},\ldots,u_j^{(q)})^T, \vv_j=(v_j^{(1)},\ldots,v_j^{(q)})^T$ is the minimizer of
$$\min_{\vu_j,\vv_j\in (L_2(P))^q} E[\|\vY_i-\vu_j(T_i)-\vv_j(T_i)X_{ij}\|^2|T_i].$$

As a direct extension of what is discussed in Section 2.1 of \cite{fanmadai13}, we can show that in  multivariate regression, we have
$$\beta_j-E[\|E[\vY_i|T_i]\|^2]=E\left[\frac{\|Cov(X_{ij},\vY_i|T_i)\|^2}{Var[X_{ij}|T_i]}\right],$$
where $Cov(X_{ij},\vY_i|T_i)=(Cov(X_{ij},Y_{i1}|T_i),\ldots, Cov(X_{ij},Y_{iq}|T_i))^T$. Thus $\beta_j$ is closely related to marginal conditional correlation between $X_{ij}$ and $\vY_{i}$.

Let $\calM=\{j: f_{0j}^{(l)}\neq 0 \mbox{ for at least one } l\}$ (or, if the same ordering of variables is adopted as in the previous section $\calM=\{1,\ldots,s\}$ but in this section we do not require $s$ to be fixed). To derive sure screening property, we need to assume that for the covariates associated with nonzero coefficients $(j\in\calM)$, $\beta_j$ does not vanish.
\begin{itemize}
\item[(s1)] $\min_{j\in \calM}\beta_j\ge qn^{-\kappa}$ for some $\kappa\in (0,1/2)$.
\end{itemize}
We also assume the correct rank is used.
\begin{itemize}
\item[(s2)] The matrix $(\vu_j(T_1)+\vv_j(T_1)X_{1j},\ldots,\vu_j(T_n)+\vv_j(T_n)X_{nj})$ has rank no larger than $r_j$. Let $r=\max_j{r_j}$.
\end{itemize}
Let $\vw_{ij}=\vY_{i}-\vu_j(T_i)-\vv_j(T_i)X_{ij}$. The following regularity conditions are required for uniform (over $j$) convergence properties of the spline estimator.
\begin{itemize}
\item[(s3)] For $g=u_j^{(l)}$ or $g=v_j^{(l)}$, $g$ satisfies a Lipschitz condition of order $d>1/2$. The order of the B-spline used satisfies $m\ge d+2$.
\item[(s4)] The variables $\vw_{ij}$ are uniformly (over $j$) subGaussian, that is, there exist constants $\sigma>0, L>0$ such that $E[\exp\{\|\vw_{ij}\|^2/(qL^2)\}]-1\le \sigma^2/L^2$.
\end{itemize}

Assumption (s1)  is necessary for ensuring the sure screening property, and was also assumed in \cite{fanmadai13}. Assumption (s4) makes it possible to use Bernstein's inequality in Lemma \ref{lem:w} in the Appendix to control the tail probability of $\|\vw_{ij}\|$, although it is possible to use weaker conditions on $\|\vw_{ij}\|$ to get a weaker version of Lemma \ref{lem:w}.

\begin{thm}\label{thm:screen}
Suppose (c1) and (s1)-(s4) hold, $n^{\kappa}/K^d\rightarrow 0$, and $(q+K)r\log(n)^2/(qn^{1-2\kappa})\rightarrow 0$, then
\begin{gather*}
P(\max\nolimits_{j\in \calM}|\hat{\beta}_{j}-\beta _{j}|>qn^{-\kappa }/2) \\
\leq C_{1}|\calM|\left( \exp \{-C_{2}(q+K)r\}+\exp
\{-C_{3}n/K^{2d}\}+K^{2}\exp \{-C_{4}n/K^{3}\}\right),
\end{gather*}
where $C_1,\ldots,C_4$ are positive constants.
As a corollary, if we take $\theta_n=qn^{-\kappa}/2$, we have the sure screening property
$$P(\calM\subseteq\hat{\calM})\ge 1-C_1|\calM|\left(\exp\{ -C_2(q+K)r\}+\exp\{ -C_3n/K^{2d}\}+K^2\exp\{ -C_4n/K^3\}\right),$$
with the right hand side converging to one if $\log|\calM|/\min\{(q+K)r,nK^{-2d},n/K^3\}\rightarrow 0$.
\end{thm}
\begin{rmk} We note that the assumptions on $K,q,r$ can be satisfied under various situations. For example, if $K=n^{1/(2d+1)}$ (this is the optimal choice of $K$ in classical nonparametric regression), $r$ is bounded, these assumptions are satisfied as long as $\kappa<d/(2d+1)$.
\end{rmk}
To improve the performance of screening, iterative methods can be applied \citep{fan2008sure,fanmadai13}. In our current study, we do not pursue this direction in detail due to its more computationally intensive nature, and simply use screening on our real data to reduce the number of covariates to a level that is amenable for penalized regression computationally. In our simulation studies for the performance of screening, we mainly investigate whether there is any advantage in using reduced-rank regression in the screening step.

\section{Numerical Examples}
\subsection{Simulations}
We generate data from the model
\[Y_{il}=\sum_{j=1}^pf_j^{(l)}(T_{i})X_{ij}+\epsilon_{il}, i=1,\ldots, n,\; l=1,\ldots,q.\]
More specifically, the covariates are generated from a multivariate Gaussian distribution with mean zero and covariance $Cov(X_{ij_1},X_{ij_2})=\rho^{|j_1-j_2|}$ and the index variable $T_i$ are generated from a uniform distribution on $[0,1]$. We then generate a random $q\times r$ matrix $\vA$ whose entries are i.i.d. $N(0,1)$. An $n\times 2$ matrix $\vN$ is constructed as $\vN=\vN_1+\cdots +\vN_p$ with
$$
\vN_j=\left(\begin{array}{cc}
		g_j^{(1)}(T_1)X_{1j} & g_j^{(2)}(T_1)X_{1j}\\
                    \vdots & \vdots\\
		g_j^{(1)}(T_n)X_{nj} & g_j^{(2)}(T_n)X_{nj}\\
		\end{array}\right).
$$
Finally we generate $\vY=\vN\vA^T+\vE$ where the entries of the error matrix $\vE$ are i.i.d. $N(0, \sigma^2)$. Thus $f_j^{(l)}$ are defined as (random) linear combinations of $g_j^{(1)}$ and $g_j^{(2)}$.

We set $n=100$, $q=5$ or $20$, $p=50$ or $200$, $\sigma=0.5$ and $\rho=0.3$. The nonparametric coefficient functions are specified as
$$g_1^{(1)}(t)=4\sin(2\pi t)/(2-\sin(2\pi t)), g_2^{(1)}(t)=4\exp\{5t-1\},$$
$$g_3^{(1)}(t)=2t\sin(2\pi t), g_4^{(1)}(t)=10(t-0.5)^2\exp\{-t^2\}$$
and
$$g_1^{(2)}(t)=2t\sin(2\pi t), g_2^{(2)}(t)=10(t-0.5)^2\exp\{-t^2\},$$
$$g_3^{(2)}(t)=4\sin(2\pi t)/(2-\sin(2\pi t)), g_4^{(2)}(t)=4\exp\{5t-1\},$$
and $g_j^{(l)}, j\ge 5$ are all zeros. $500$ simulated datasets are generated in each scenario.

Table \ref{tab:vs} reports the model selection results for joint variable and rank selection using either the group lasso or the group SCAD penalty. We report the percentage of times that rank of 1,2,3 are selected, and the number of nonzero coefficients identified, as well as the number of nonzero coefficients that are nonzero in the true model. It is seen that the correct rank is selected most of the time. In terms of variable selection, the nonzero coefficients in the true model are always selected with satisfactory false positive rate. Model selection using the SCAD penalty is slightly better than the lasso penalty.

\begin{center}
\textit{Insert Table \ref{tab:vs} here }
\end{center}

In Figures \ref{fig:p50q5}-\ref{fig:p200q20}, mean squared errors (MSE) $\|\hat{f}_j-f_j\|^2$, for $j=1,\ldots,5$, as well as the total errors $\sum_{i=1}^p\|\hat{f}_j-f_j\|^2$, are shown for the oracle estimator (ORA, the true rank and only the true nonzero coefficients are used in modeling fitting without penalty), our estimator with either the lasso (LAS) or SCAD penalty (SCAD), and the estimator that only uses lasso or SCAD penalties for variable selection without rank constraint (LAS-FR and SCAD-FR). We also fitted linear models to the data but since the true model is nonlinear the errors for the linear models are much larger and thus not shown here. In terms of estimation errors, SCAD-penalized estimators perform better than lasso-penalized estimators, and using rank constraint generally helps to improve performance.

\begin{center}
\textit{Insert Figures \ref{fig:p50q5}-\ref{fig:p200q20} here }
\end{center}

Now we study the performance of the screening procedure and investigate whether there is any advantage of using rank constraint in the screening step. We first use a similar setup as before with $p=1000$ and $q=20$. Again $500$ datasets are generated. The rank in each marginal regression problem is selected based on 5-fold cross-validation error. We compare the screening procedure with rank constraint and that without rank constraint in terms of the rankings assumed by the first four covariates after the covariates are sorted by $\hat{\beta}_j$. The first two rows of Figure \ref{fig:screen1} show the histograms of rankings for the first four covariates. The histograms obtained from reduced-rank screening and full-rank screening are visually almost the same. In the third row of Figure \ref{fig:screen1}, the empirical distribution of the rankings for the four covariates are compared, with the black curve being the distribution of rankings for reduced-rank screening and the red curve being the distribution of rankings for full-rank screening. These plots clearly show that there is no advantage of using reduced-rank screening in this setting.

We think the reason that reduced-rank screening has no advantages in screening under the current simulation setting is that when the estimation is not sufficiently hard, the additional degree of parsimony provided by reduced-rank screening does not result in significantly better estimation of $\beta_j$. Thus we increase the standard deviation of the noise from $0.5$ to $1$ and $2$ and performed the simulations with these two larger noise levels with results reported in Figures \ref{fig:screen2} and \ref{fig:screen3}. We see that with larger variance of the noise that makes the estimation problem more difficult, the advantages of reduced-rank screening start to appear. This can be seen from the observation that the red curve is more frequently below the black curve (in particular for the 1st and 3rd covariates in Figure \ref{fig:screen3}). We think the situation with very low signal to noise ratio may actually be more realistic in high-dimensional real data analysis and thus reduced-rank screening can help, although it seems to only have a relatively small effect. 

\begin{center}
\textit{Insert Figures \ref{fig:screen1}-\ref{fig:screen3} here }
\end{center}

\subsection{Framingham Heart Study}

We use 15 continuous variables as the responses which are described in Table %
\ref{tab:response}. The index variable is the level of sedentary activity in
term of hours per day. Each SNP has three possible allele combinations
denoted as $-1$, $0$, $1$. For details on genotyping, see
\url{http://www.ncbi.nlm.nih.gov/projects/gap/cgi-bin/study.cgi?study_id=phs000007.v20.p8}. { We then code the three genotype categories using two
dummy variables as $(0,0)$, $(0,1)$ and $(1,0)$. } For illustration purposes, we focus on identifying important
SNPs located at chromosome 4.

\begin{center}
\textit{Insert Table \ref{tab:response} here }
\end{center}

In the first step, we conduct the independence screening procedure. As a
result, 250 SNPs are selected. In the second step, we perform the proposed
penalization estimation with rank constraint in the varying coefficient
model by using the selected SNPs. Note that with 250 SNPs and $K=5$, the
effective dimension of the regression is $2500$ using the above variable
coding method. Thus to go beyond $250$ SNPs, more efficient computational
approach for penalized regression needs to be developed in the future. We
compare our procedure with several others, including one similar to our
approach except that group lasso penalty is used (LASSO), two methods that
use penalized variable selection without rank constraint (LASSO-FR and
SCAD-FR). The 5-fold cross-validation errors of different methods (with
tuning parameter chosen by 5-fold cross-validation), as well as the number
of SNPs selected, are reported in Table \ref{tab:real}. We see that using
rank constraint indeed helps reduce prediction error. The errors from the
two kinds of penalties are similar with the SCAD penalty slightly better in
this example. Finally, nonlinear methods identify a smaller number of SNPs
than linear methods. The estimated coefficients with the SCAD penalty are
shown in Figure \ref{fig:real}.


\begin{center}
\textit{Insert Table \ref{tab:real} here }
\end{center}

\begin{center}
\textit{Insert Figure \ref{fig:real} here }
\end{center}

\section{Conclusion and Discussion}
In this paper, we proposed a dimension reduction and variable selection method in multivariate varying coefficient models, in which effects of the covariates are allowed to change with another variable. As a result, it provides a more flexible approach than parametric multivariate regression. Moreover, we established the convergence rate of the nonparametric coefficient estimators as well as the variable selection consistency. A screening procedure is also proposed for the cases with ultrahigh dimensional predictors. Our numerical results demonstrate the advantages of combining rank constraint, variable selection, and the varying coefficient structure. The tuning parameters $r$ and $\lambda$ were selected by using cross-validation, although theoretical investigations concerning its optimality properties seem challenging.

Instead of using constraint $rank(\vC)\le r$, an alternative way is to penalize the nuclear norm of $\vC$ as in \cite{negahban11,koltchinskii11}. As shown in \cite{bunea11} for parametric models, nuclear norm penalized estimator has estimation properties similar to those of rank constrained estimator, although it often results in less parsimonious models (with a larger rank selected). Entirely different computational algorithms also need to be developed if nuclear norm penalty is added. It is outside the scope of the current paper to compare the two estimators numerically.

The proposed method has a wide range of data applications, and it is particularly useful to identify variables when their effects can change with another variable in high-dimensional cases, such as gene-environment interactions in genome-wide association studies (GWAS). The method can be straightforwardly extended to other structural models such as additive models and single-index models. In our method, we require the response variables are continuous. In real data applications, however, discrete response variables may occur, such as disease status. Thus how to incorporate both continuous and discrete responses in the dimension reduction and variable selection procedure can be a future topic. Moreover, the FHS dataset is a continuing project containing longitudinal observations, so extending the proposed method to longitudinal data settings is also of our interest, which needs further investigation.

\section*{Appendix A: Proof for Theorems \ref{thm:rates} and \ref{thm:vs}}
We first introduce some notations and additional definitions. In our proofs, $C, C_1,C_2,\ldots,$ denote generic positive constants that might assume different values at different places. 
Recall the definition of $\vC_0$ just before the statement of Theorem \ref{thm:rates}. By our assumptions $\vC_0$ can be written as $\vB_0\vA_0^T$ with $\vB_0=(\vB_{00}^T,\ldots,\vB_{0p}^T)^T$ with $\vB_{0j}$ being $K\times r$ matrices and $\vA_0$ a $q\times r$ matrix. Let $\vZ_{S}=(\vZ_0,\ldots,\vZ_{s})$ be the $n\times (s+1)K$ submatrix of $\vZ$ containing the columns corresponding to nonzero predictors. More generally, we will use subscript $S$ to denote other subvectors/submatrices associated with nonzero predictors. We assume the constraint $rank(\vC)\le r$ is used (that is the constraint is correctly specified). Note that in practice $r$ is chosen by five-fold cross-validation.

We prove the two theorems together and the proof consists of two steps. Roughly speaking, we first show that the ``oracle estimator" which assumes knowledge of zero blocks $\vC_j, j>s$ achieves the convergence rate stated in Theorem \ref{thm:rates} and then we show that this oracle estimator is actually a local partial optimum of (\ref{eqn:min}), which will complete the proof.

Formally, we define the oracle estimator as
$$\tilde\vC_S=\argmin_{\vC_S:rank(\vC_S)\le r}\|\vY-\vZ_S\vC_S\|^2+n\sum_{j=1}^s p_\lambda(\|\vC_j\|).$$
For simplicity of notation below, we denote the objective functional above as $Q_n(\vB,\vA)$ when we write $\vC_S=\vB\vA^T$.

Although in computation, we used the parameterization $\vC=\vB\vA^T$ with $\vA^T\vA=\vI$. This orthonormality condition is inconvenient in theoretical derivations for our first part of the proof and would require studying Stiefel manifold structure as in \cite{chenhuang12}. Thus we instead use the parameterization adopted in \cite{kunchen12}. More specifically, since $\vA$ has rank $r$, there is an $r\times r$ submatrix $\vA'$ of $\vA$ that is invertible and replacing $\vB$ by $\vB(\vA')^T$ and $\vA$ by $\vA(\vA')^{-1}$, we can assume there are $r$ rows of $\vA$ such that the submatrix consisting of these $r$ rows is an identity matrix. Without loss of generality we assume the submatrix of $\vA_0$ consisting of its first $r$ rows is the identity matrix.  For ease of notation, we still write $\vC_S=\vB\vA^T$ for a $(s+1)K\times r$ matrix $\vB$ and a $q\times r$ matrix $\vA$ where now $\vA$ is no longer an orthogonal matrix. The parameter space we consider in a neighborhood of $(\vB_0,\vA_0)$ is $\Omega=\{\vB\vA^T: \vB \mbox{ is a } (s+1)K\times r \mbox{  matrix}, \vA \mbox{  is an } q\times r  \mbox{  matrix, and the first } r \mbox{  rows of } $\vA$ \mbox{ is the identity matrix}\}$.

In the first part of the proof, we only consider the oracle estimator. We omit the subscript $S$ in the following for simplicity.
Now let $\vB=\vB_0+\vU$ and $\vA=\vA_0+\vV$, with the first $r$ rows of $\vV$ being zero. Denote $\vD(\vU,\vV)=\vB\vA^T-\vB_0\vA_0^T=\vU\vA_0^T+\vB_0\vV^T+\vU\vV^T$. We will show that for any $\epsilon>0$, there exists a large enough constant $C>0$ such that
\begin{equation}\label{eqn:eps}
P\left( \inf_{(\vU,\vV):\|\vD(\vU,\vV)\|=C\gamma_n} Q_n(\vB,\vA)-Q_n(\vB_0,\vA_0)>0\right)\ge 1-\epsilon,
\end{equation}
where $\gamma_n=\sqrt{K/n}(\sqrt{(q+K)r}+\sqrt{nq}/K^d)$.
This will imply that there is a local minimizer of $Q_n$, $(\tilde{\vB},\tilde{\vA})$ with $\|\tilde{\vB}\tilde{\vA}^T-\vB_0\vA_0^T\|=O_p(\gamma_n)$. Using the approximation properties of splines, the convergence rate (\ref{eqn:rates2}) for the oracle estimator immediately follows from (\ref{eqn:rates1}).

To show (\ref{eqn:eps}), we have
\begin{eqnarray*}
&&Q_n(\vB,\vA)-Q_n(\vB_0,\vA_0)\\
&=&\|\vY-\vZ\vB_0\vA_0^T-\vZ\vD(\vU,\vV)\|^2-\|\vY-\vZ\vB_0\vA_0^T\|^2+n\sum_{j=1}^s p_\lambda(\|\vB_j\vA^T\|)-n\sum_{j=1}^sp_\lambda(\|\vB_{0j}\vA_0^T\|).
\end{eqnarray*}
Since $\|\vB_{0j}\vA_0^T\|$ is of order $\sqrt{qK}$ and $\|\vB_j\vA^T-\vB_{0j}\vA_0^T\|\le \|\vD(\vU,\vV)\|=C\gamma_n=o(\sqrt{qK})$, we have $\sum_{j=1}^s p_\lambda(\|\vB_j\vA^T\|)-\sum_{j=1}^sp_\lambda(\|\vB_{0j}\vA_0^T\|)=0$, by the assumption on $\lambda$ and the property that that $p_\lambda(x)$ is a constant when $x>a\lambda$. Thus the above can be continued as
\begin{eqnarray*}
&&Q_n(\vB,\vA)-Q_n(\vB_0,\vA_0)\\
&=&\|\vZ\vD(\vU,\vV)\|^2-2\langle \vZ\vD(\vU,\vV), \vY-\vZ\vB_0\vA_0^T\rangle\\
&=&\|\vZ\vD(\vU,\vV)\|^2-2\langle \vZ\vD(\vU,\vV), \vE\rangle -2\langle \vZ\vD(\vU,\vV),\vM-\vZ\vB_0\vA_0^T\rangle,\\
\end{eqnarray*}
where $\langle \vF,\vG\rangle={\rm tr}(\vF\vG^T)$ for any two matrices $\vF,\vG$. By Lemma A.1 of \cite{huang04}, the eigenvalues of $\vZ^T\vZ$ are of order $n/K$. Thus
\begin{equation}\label{eqn:combine1}
\|\vZ\vD(\vU,\vV)\|^2\sim (n/K)\|\vD(\vU,\vV)\|^2.
\end{equation}
We also get from the approximation error
\begin{equation}\label{eqn:combine2}
\|\vM-\vZ\vB_0\vA_0^T\|=O_p(\sqrt{nq}K^{-d}).
\end{equation}
Furthermore, $\langle \vZ\vD(\vU,\vV), \vE\rangle=\langle \vZ\vD(\vU,\vV), \vP_\vZ\vE\rangle$ where $\vP_\vZ=\vZ(\vZ^T\vZ)^{-1}\vZ^T$. We now use the duality inequality $ \langle \vZ\vD(\vU,\vV), \vP_\vZ\vE\rangle\le d_1(\vP_\vZ\vE)\|\vZ\vD(\vU,\vV)\|_*$, where $\|.\|_*$ is the nuclear norm (sum of singular values). We have
\begin{equation}\label{eqn:combine3}
d_1(\vP_\vZ\vE)=O_p(\sqrt{q+K}),
\end{equation}
by Lemma 3 in \cite{bunea11} and
\begin{equation}\label{eqn:combine4}
\|\vZ\vD(\vU,\vV)\|_*\le \sqrt{2r}\|\vZ\vD(\vU,\vV)\|,
\end{equation}
since $rank(\vD(\vU,\vV))\le 2r$. Combining (\ref{eqn:combine1})-(\ref{eqn:combine4}), we get that
$Q_n(\vB,\vA)-Q_n(\vB_0,\vA_0)$ is bounded below by a constant multiple of
$$\frac{n}{K}\|\vD(\vU,\vV)\|^2- \sqrt{n/K}\|\vD(\vU,\vV)\|\sqrt{(q+K)r}-\sqrt{n/K}\|\vD(\vU,\vV)\|\sqrt{nq}K^{-d},$$
which is easily seen to be positive if $\|\vD(\vU,\vV)\|=C\gamma_n$ with $C$ large enough.

Now, we come to the second part of the proof where we show that the local minimizer $\tilde\vC_S=\tilde\vB\tilde\vA^T$ defined above in the $\gamma_n$-neighborhood of $\vC_{0S}$ is a local partial optimum of the original problem (\ref{eqn:min}) (without the information regarding zero rows of $\vB$), in the sense that if we set $\hat{\vA}=\tilde{\vA}$, and $\hat{\vB}_j=\tilde{\vB}_j, j\le s$, $\hat{\vB}_j=\vnull, j>s$, then $\hat{\vA},\hat{\vB}$ is a local partial optimum of (\ref{eqn:min}). In this part, we revert back to the parameterization with $\vA^T\vA=\vI$ due to that $\vA^T\vA=\vI$ can be used to simplify some expressions below. Note that since $(\tilde{\vB}, \tilde{\vA}^T)$ is a local minimizer of a functional that depends only on $\tilde\vB\tilde\vA^T$, any parameterization will not change the fact that it is a local minimizer.

That $\hat{\vA}$ is a local minimizer of (\ref{eqn:min}) for fixed $\hat{\vB}$ trivially follows from the definition of the local minimizer (a local minimizer is a local partial optimum). The proof of that $\hat{\vB}$ is a local minimizer of (\ref{eqn:min}) for fixed $\hat{\vA}$ is based on the following claim, which is a direct extension of Theorem 1 in \cite{fan2009non} to group penalty (but here we specialized it to only the SCAD penalty for specificity). A similar second-order sufficiency was also used in \cite{kim2008smoothly} in linear models (see the proof of their Theorem 1). Thus the proof of the following claim is omitted.
\begin{clm}\label{clm}
$\vB\in R^{(p+1)K\times r}$ is a local minimizer of (\ref{eqn:min}) for fixed $\vA$ if
\begin{eqnarray}
&&2\vZ_j^T(\vY\vA-\vZ\vB)=0 \mbox{ and } \|\vB_j\vA^T\|\ge a\lambda, \mbox{ for } j\le s; \label{eqn:clm1}\\
&& \max_{j> s}\|2\vZ_j^T(\vY\vA-\vZ\vB)\|< n\lambda \mbox{ and } \|\vB_j\vA^T\|< \lambda \mbox{ for } j>s.\label{eqn:clm2}
\end{eqnarray}
\end{clm}
We will finish our proof by verifying the above conditions. The first part of (\ref{eqn:clm1}) follows directly from the first order optimality condition by the definition of the oracle estimator. Also $\|\hat{\vB}_j\hat{\vA}^T\|\ge \|{\vB}_{0j}\vA_0^T\|-\|\hat{\vC}_j-\vC_{0j}\|$ with $\|\vB_{0j}\vA_0^T\|\ge C\sqrt{Kq}$ and $\|\hat{\vC}_j-\vC_{0j}\|=O(\gamma_n)=o(\sqrt{Kq})$ and thus $\|\hat{\vB}_j\hat{\vA}^T\|\ge C\sqrt{Kq}\ge a\lambda$ with probability approaching one. Thus (\ref{eqn:clm1}) is verified.

For (\ref{eqn:clm2}), that $\|\hat{\vB}_j\hat{\vA}^T\|< \lambda$ is trivially satisfied since $\|\hat{\vB}_j\hat{\vA}^T\|=0$. For the first part of (\ref{eqn:clm2}), we have
\begin{eqnarray*}
&&\vZ_j^T(\vY\hat{\vA}-\vZ\hat{\vB})\\
&=&\vZ_j^T\vE\vA_0+\vZ_j^T(\vM-\vZ\vB_0\vA_0^T)\vA_0+\vZ_j^T\vY(\hat{\vA}-\vA_0)-\vZ_j^T\vZ(\hat{\vB}-\vB_0)
\end{eqnarray*}
Similar to Lemma 2 of \cite{huang10}, when $K\log(pK)/n\rightarrow 0$, we have $\max_{j>s}\|\vZ_j^T\vE\vA_0\|\le\max_{j>s}\|\vZ_j^T\vE\|=O_p(\sqrt{nq}\sqrt{\log(pK)})=o(n\lambda)$. Furthermore, using $\|\vZ_j^T(\vM-\vZ\vB_0\vA_0^T)\vA_0\|=O_p(\sqrt{n/K}\sqrt{nq}K^{-d})=o(n\lambda)$, $\|\vZ_j^T\vY(\hat{\vA}-\vA_0)\|=O_p(\sqrt{n/K}\sqrt{nq}\gamma_n)=o(n\lambda)$, and $\|\vZ_j^T\vZ(\hat{\vB}-\vB_0)\|=(n/K)\gamma_n=o(n\lambda)$, we verified that (\ref{eqn:clm2}) is satisfied with probability approaching one.

\section*{Appendix B: Proof for Theorem \ref{thm:screen}}

For simplicity of notation in the proof we assume $r_j\equiv r$. As usually done in spline estimation to bridge true functions and estimated spline functions, we define the population counterpart of $\hat{\beta}_j$ using B-spline basis. Let $\vH_{0j}$ be the minimizer of
$$\min_{ rank(\vH_j)\le r}E[\|\vY_i-\bar\vZ_{ij}\vH_j\|^2],$$
and define
$$\tilde{\beta}_j=E[\|\bar\vZ_{ij}\vH_{0j}\|^2],$$
where $\bar\vZ_{ij}$ is the $i$-th row of $\bar\vZ_j$. We also let $\tilde{\beta}_{nj}=\frac{1}{n}\sum_{i=1}^n\|\bar\vZ_{ij}\vH_{0j}\|^2$.

We first state several lemmas that will be used in the proof of the theorem.

\begin{lem}\label{lem:eigen}
There exist positive constants $C_1,C_2$ such that all the eigenvalues of $E\bar\vZ_j^T\bar\vZ_j/n$ are inside the interval $[C_1/K,C_2/K]$ for all $j$, and all the eigenvalues of $\bar\vZ_j^T\bar\vZ_j/n$ are inside the interval $[C_1/(2K),2C_2/K]$ with probability at least $1-C_3K^2\exp\{-C_4n/K^3\}$ for all $j$.
\end{lem}
\textbf{Proof of Lemma \ref{lem:eigen}.} This result is just a restatement of Lemma 7 of \cite{fanmadai13}. \hfill $\Box$

\begin{lem}\label{lem:covering}
Let $\calH_j(A)=\{\vH_j\in R^{2K\times q}:  rank(\vH_j)\le r, \|\bar\vZ_j(\vH_j-\vH_{0j})\|\le A\}$ and $\calH'_j(A)=\{\vH_j\in R^{2K\times q}:  rank(\vH_j)\le r, \|\vH_j-\vH_{0j}\|\le A\}$. Then there is a constant $C>0$ such that for any $A>0$
$$P(\calH_j(A)\subseteq \calH_j'(C\sqrt{K/n}A))\ge 1-C_1K^2\exp\{-C_2n/K^3\}. $$
In particular, this implies that
$$H(\epsilon,\calH_j(A),\|.\|)\le H(\epsilon,\calH_j'(C\sqrt{K/n}A),\|.\|)$$
with probability at least $1-C_1K^2\exp\{-C_2n/K^3\}$, where $H(\epsilon,\calF,d)$ is the $\epsilon$-entropy number (logarithm of covering  number) of $\calF$ with metric $d$.
\end{lem}
\textbf{Proof of Lemma \ref{lem:covering}.} This is a corollary of Lemma \ref{lem:eigen} since $\lambda_{\min}(\bar\vZ_j^T\bar\vZ_j)\|\vH_j-\vH_{0j}\|^2\le\|\bar\vZ_j(\vH_j-\vH_{0j})\|^2\le \lambda_{\max}(\bar\vZ_j^T\bar\vZ_j)\|\vH_j-\vH_{0j}\|^2$.

\begin{lem}\label{lem:covering2}
With the set $\calH_j'(A)$ as defined above, its entropy can be bounded by
$$H(\epsilon,\calH_j'(A),\|.\|)\le C(q+K)r\log((A+1)\sqrt{r}/\epsilon).$$
\end{lem}
\textbf{Proof of Lemma \ref{lem:covering2}.}
Due to the rank assumption, we can write $\vH_{0j}=\vB_{0j}^T\vA_{0j}^T$ with $\vB_{0j}\in R^{2K\times r}$, $\vA_{0j}\in R^{q\times r}$ and $\vA_{0j}^T\vA_{0j}=\vI$. Similarly, any $\vH\in \calH_j'(A)$ can be written as $\vH=\vB\vA^T$ with $\vB\in R^{2K\times r}, \|\vB-\vB_{0j}\|=\|\vH-\vH_{0j}\|\le A$ and $\vA\in \calO^{q\times r}=\{\vA\in R^{q\times r}: \vA^T\vA=\vI\}$. Note that $\|\vA\|^2=r$ for all $\vA\in \calO^{q\times r}$.  Let $\calB=\{\vB_1,\ldots,\vB_{N_1}\}$ be an $\epsilon/2$-covering of $\{\vB: \|\vB-\vB_{0j}\|\le A$ in $R^{2K\times r}$ and $\calA=\{\vA_1,\ldots,\vA_{N_2}\}$ be a $\epsilon/(2\|\vH_{0j}\|+2A)$-covering of $\calO^{q\times r}$ in $\{\vA: \|\vA\|\le \sqrt{r}\}$ (that is $\vA_k, k=1,\ldots, N_2$ are not necessarily elements of $\calO^{q\times r}$). By Lemma 2.5 of \cite{geer00}, we have $N_1\le (8A/\epsilon+2)^{2Kr}$ and $N_2\le (8\sqrt{r}(\|\vH_{0j}\|+A)/\epsilon+2)^{qr}$. We can further project elements in $\calA$ on $\calO^{q\times r}$ to get a covering with elements inside $\calO^{q\times r}$.

Since
$\|\vB\vA^T-\vB_k\vA_{k'}^T\|\le \|\vB-\vB_k\|+\|\vB_k\|\cdot\|\vA-\vA_{k'}\|$, it is easy to see that $\{\vH: \vH=\vB_k\vA_{k'}^T, \mbox{ for some }\vB_k\in\calB,\vA_{k'}\in\calA\}$ is an $\epsilon$-covering of $\calH'_j(A)$ with entropy number  bounded by $C(q+K)r\log((A+1)\sqrt{r}/\epsilon)$.
\hfill $\Box$

\begin{lem}\label{lem:int}
There is a constant $C$ such that for any $\delta>0$,
$$\int_0^\delta\log^{1/2}(\frac{(\delta+1)\sqrt{r}}{\epsilon})d\epsilon\le C \delta (\log(\frac{1}{\delta}) \vee \log(r)\vee 1)$$
\end{lem}
\textbf{Proof of Lemma \ref{lem:int}.} Trivially we have $\int_0^\delta \log^{1/2}\sqrt{r}d\epsilon=\delta\log^{1/2}(r)\le C\delta\log(r)$.

For $\delta>1/20$, we have
\begin{eqnarray*}
&&\int_0^\delta\log^{1/2}(\frac{\delta+1}{\epsilon})d\epsilon\\
&\le & \int_0^\delta \log^{1/2}(\frac{21\delta}{\epsilon})d\epsilon\\
&=&\delta \int_1^\infty \frac{\log^{1/2}(21y)}{y^2}dy\\
&\le & C\delta,
\end{eqnarray*}
as long as $C\ge \int_1^\infty \frac{\log^{1/2}(21y)}{y^2}dy$.
For $\delta\le 1/20$, we have
\begin{eqnarray*}
&&\int_0^\delta\log^{1/2}(\frac{\delta+1}{\epsilon})d\epsilon\\
&\le & \int_0^\delta \log^{1/2}(\frac{2}{\epsilon})d\epsilon\\
&\le & \int_{1/\delta}^\infty\frac{\log^{1/2}(2y)}{y^2}dy.
\end{eqnarray*}
For ease of notation, let $a=1/\delta\ge 20$. Define the function
$$f(a)=\int_a^\infty\frac{\log^{1/2}(2y)}{y^2}dy-\frac{1}{a}\log(a).$$
Direct calculation shows $f'(a)=-\frac{\log^{1/2}(2a)}{a^2}-\frac{1-\log(a)}{a^2}=\frac{\log(a)-\log^{1/2}(2a)-1}{a^2}>0$ when $a\ge 20$, and that $\lim_{a\rightarrow\infty}f(a)=0$ and thus $f(a)<0$ for $a\ge 20$ and this implies
\begin{eqnarray*}
&& \int_{1/\delta}^\infty\frac{\log^{1/2}(2y)}{y^2}dy\\
&\le & \delta\log(\frac{1}{\delta}).
\end{eqnarray*}
\hfill $\Box$

\begin{lem}\label{lem:w}
$$P(\|\vW_j\|^2>2nq\sigma^2)\le C_1 \exp\{-C_2n\},$$
where $\vW_j$ are the $n\times q$ matrix with rows $\vw_{ij}$.
\end{lem}
\textbf{Proof of Lemma \ref{lem:w}.} Since $E[\exp\{\|\vw_{ij}\|^2/(qL^2)\}]-1\le \sigma^2/L^2$ and using Taylor's expansion $e^x=\sum_{i=0}^\infty x^i/i!$, it is easy to see that $E\|\vw_{ij}\|^2\le q\sigma^2$ and $E[\|\vw_{ij}\|^{2m}]\le (\frac{m!}{2})(qL^2)^{m-2}(2q^2L^2\sigma^2), m\ge 2$. Applying Bernstein's inequality (Lemma 8.6 of \cite{geer00}), we get
$$P(|\frac{1}{n}\sum_i\|\vw_{ij}\|^2-E\|\vw_{ij}\|^2|>qa)\le 2\exp\{-na^2/(aL^2+2L^2\sigma^2)\}.$$
Taking $a=\sigma^2$, this implies
$$P(\frac{1}{n}\sum_i\|\vw_{ij}\|^2>2q\sigma^2)\le 2\exp\{-n\sigma^2/(3L^2)\}.$$
\hfill $\Box$

\textbf{Proof of Theorem \ref{thm:screen}.} First using the spline approximation property we have
\begin{equation}\label{eqn:comb1}
|\beta_j-\tilde{\beta}_j|=O(qK^{-d}).
\end{equation}

Next, since $\|\bar\vZ_{ij}\vH_{0j}\|^2=O({q})$, using Bernstein's inequality, we have
\begin{equation}\label{eqn:comb2}
P(|\tilde{\beta}_{nj}-\tilde{\beta}_j|>C_1{q}K^{-d})\le 2 \exp\{-C_2 n/K^{2d}\}.
\end{equation}

Furthermore, we have
\begin{eqnarray}\label{eqn:diff}
|\hat{\beta}_j-\tilde{\beta}_{nj}|&=&\frac{1}{n}\sum_{i=1}^n\left|\|\bar\vZ_{ij}\hat{\vH}_j\|^2-\|\bar\vZ_{ij}{\vH}_{0j}\|^2\right|\nonumber\\
&\le &\frac{1}{n}\sum_{i=1}^n\|\bar\vZ_{ij}(\hat{\vH}_j-\vH_{0j})\|\cdot(\|\bar\vZ_{ij}\hat{\vH}_{j}\|+ \|\bar\vZ_{ij}{\vH}_{0j}\|)\nonumber\\
&\le &(\frac{1}{n}\sum_{i=1}^n\|\bar\vZ_{ij}(\hat{\vH}_j-\vH_{0j})\|^2)^{1/2}\cdot(\frac{2}{n}\sum_{i=1}^n\|\bar\vZ_{ij}\hat{\vH}_{j}\|^2+ \frac{2}{n}\sum_{i=1}^n\|\bar\vZ_{ij}{\vH}_{0j}\|^2)^{1/2}\nonumber\\
&=&(C/n)\|\bar\vZ_j(\hat{\vH}_j-\vH_{0j})\|\cdot(\|\bar\vZ_j\hat{\vH}_j\|+\|\bar\vZ_j{\vH}_{0j}\|).
\end{eqnarray}

Since $\hat{\vH}_j$ is a minimizer of $\|\vY-\bar\vZ_j\vH_j\|^2$, using $\|\vY-\bar\vZ_j\hat{\vH}_j\|^2\le \|\vY-\bar\vZ_j\vH_{0j}\|^2$, some simple algebra shows
$$\|\bar\vZ_j(\hat{\vH}_j-\vH_{0j})\|^2\le 2\langle \vW_j, \bar\vZ_j(\hat{\vH}_j-\vH_{0j})\rangle$$
which implies
$$\|\bar\vZ_j(\hat{\vH}_j-\vH_{0j})\|\le 2\|\vW_j\|,$$
and thus
\begin{eqnarray*}
&&P(\|\bar\vZ_j(\hat{\vH}_j-\vH_{0j})\|\ge \delta)\\
&\le & P(\|\bar\vZ_j(\hat{\vH}_j-\vH_{0j})\|\ge \delta,\|\vW_j\|^2\le 2nq\sigma^2)+P(\|\vW_j\|^2 > 2nq\sigma^2)\\
&\le & P(\|\bar\vZ_j(\hat{\vH}_j-\vH_{0j})\|\ge \delta,\hat{\vH}_j\in \calH_j(\sqrt{8nq\sigma^2}), \|\vW_j\|^2\le 2nq\sigma^2)+P(\|\vW_j\|^2 > 2nq\sigma^2)
\end{eqnarray*}
where $\calH_j(C)=\{\vH_j\in R^{2K\times q}:  rank(\vH_j)\le r, \|\bar\vZ_j(\vH_j-\vH_{0j})\|\le C\}$.
By Lemma \ref{lem:covering}, $\calH_j(\sqrt{8nq\sigma^2})$ can be replaced  by $\calH_j'(\sqrt{CKq})$ to get
\begin{eqnarray*}
&&P(\|\bar\vZ_j(\hat{\vH}_j-\vH_{0j})\|\ge \delta)\\
&\le & P(\|\bar\vZ_j(\hat{\vH}_j-\vH_{0j})\|\ge \delta,\hat{\vH}_j\in \calH_j'(\sqrt{CKq}), \|\vW_j\|^2\le 2nq\sigma^2)\\
&&\;\;\;\;\;+P(\|\vW_j\|^2 > 2nq\sigma^2)+C_3K^2\exp\{-C_4n/K^3\}.
\end{eqnarray*}
Using Lemmas \ref{lem:covering2} and \ref{lem:int}, we have
\begin{eqnarray*}
\int_0^\delta H^{1/2}(\epsilon,\calH_j'(\delta),\|.\|)d\epsilon&\le& C\int_0^\delta \sqrt{(q+K)r}\log^{1/2}(\frac{(\delta+1)\sqrt{r}}{\epsilon})d\epsilon\\
&\le&C\sqrt{(q+K)r}\delta (\log(\frac{1}{\delta})\vee\log(r)\vee 1).
\end{eqnarray*}

The rest follows the proof of Theorem 9.1 in \cite{geer00} which is easily seen to be valid even in multivariate regression. To apply that theorem, we need to ensure we choose $\delta$ such that $\sqrt{n}\delta^2\ge C\sqrt{(K+q)r}\delta(\log(1/\delta)\vee\log(r)\vee 1)$, and it is easy to verify that we can choose $\delta=C\sqrt{\frac{(q+K)r}{n}}\left(\log(\sqrt{\frac{n}{q+K}})\vee\log(r)\vee 1\right)$. Thus we have
\begin{eqnarray*}
&&P(\|\bar\vZ_j(\hat{\vH}_j-\vH_{0j})\|\ge C\sqrt{(q+K)r}\left(\log\left(\sqrt{\frac{n}{q+K}}\right)\vee\log(r)\vee 1\right))\\
&\le & C_1\exp\left\{-C_2(q+K)r\left(\log(\sqrt{\frac{n}{q+K}})\vee\log(r)\vee 1\right)\right\}\\
&&\;\;\;\;\;\;\;\;\;\;\;\;+P(\|\vW_j\|^2 > 2nq\sigma^2)+C_3K^2\exp\{-C_4n/K^3\}\\
&\le &C_1\exp\{-C_2(q+K)r\}+C_3K^2\exp\{-C_4n/K^3\}
\end{eqnarray*}

Thus (\ref{eqn:diff}) can be continued as
\begin{eqnarray}\label{eqn:diff2}
&&|\hat{\beta}_j-\tilde{\beta}_{nj}|\nonumber\\
&\le&(C/n)\|\bar\vZ_j(\hat{\vH}_j-\vH_{0j})\|\cdot(\|\bar\vZ_j\hat{\vH}_j\|+\|\bar\vZ_j{\vH}_{0j}\|)\nonumber\\
&\le& (C/n)\sqrt{(q+K)r}\log(n)\cdot(\sqrt{nq}+\sqrt{(q+K)r}\log(n))
\end{eqnarray}
with probability at least $1-C_1\exp\{-C_2(q+K)r\}-C_3K^2\exp\{-C_4n/K^3\}$. The theorem is proved by combining (\ref{eqn:comb1}), (\ref{eqn:comb2}), (\ref{eqn:diff2}), and using the union bound over $j\in \calM$.
\hfill $\Box$

\bibliographystyle{jasa}
\bibliography{papers,books}

\clearpage\pagebreak \newpage \thispagestyle{empty}

\begin{table}[tp]
\caption{Rank and variable selection results for both the lasso and SCAD penalized estimators. }
\vspace{0.1in}
\label{tab:vs}\centering
\begin{tabular}{ccccccc}
\hline \multicolumn{2}{c}{}&{\% $\hat{r}$=1}&{\% $\hat{r}$=2}&{\% $\hat{r}$=3}&{\# nonzero}&{\# nonzero correct}\\

\hline
 {$p=50,q=5$} & LASSO & $ 0 $ & $ 92 $ & $ 4 $ &  $ 4.56 ( 0.812 )$ & $4 ( 0 )$ \\
& SCAD & $ 0 $ & $ 93 $ & $ 7 $ & $ 4.41 ( 0.606 )$ & $4 ( 0 )$ \\\\

 {$p=200,q=5$} & LASSO  & $ 0 $ & $ 83 $ & $ 10 $ & $ 4.53 ( 0.735 )$ & $4 ( 0 )$ \\
 &SCAD & $ 0 $ & $ 86 $ & $ 12 $& $ 4.35 ( 0.626 )$ & $4 ( 0 )$ \\\\

 {$p=50,q=20$} & LASSO  & $ 0 $ & $ 66 $ & $ 22 $ & $ 5.76 ( 1.756 )$ & $4 ( 0 )$ \\
& SCAD & $ 0 $ & $82 $ & $ 17 $  & $ 5.46 ( 1.580 )$ & $4 ( 0 )$ \\\\

 {$p=200,q=20$} & LASSO  & $ 0 $ & $ 56 $ & $ 32 $ & $ 6.52 ( 2.287 )$ & $4 ( 0 )$ \\
& SCAD & $ 0 $ & $ 73 $ & $ 22$ & $ 6.23 ( 2.140 )$ & $4 ( 0 )$ \\
\hline
\end{tabular}
\end{table}

\begin{table}[tp]
\caption{The response variables we use for the FHS data. }\vspace{0.1in} \centering%
\begin{tabular}{c|c}
\hline
variable name & description \\ \hline
weight & weight (to nearest pound) \\
height &  height (in inches to next lower 1/4 inch) \\
bi.deltoid.girth & bi-deltoid girth (inches with 2 decimals) \\
right.arm.girth.upper.third & right arm girth-upper third (inches with 2
decimals) \\
waist.girth & waist girth (inches with 2 decimals) \\
hip.girth & hip girth (inches with 2 decimals) \\
thigh.girth & thigh girth (inches with 2 decimals) \\
systolic.blood.pressure & systolic blood pressure-nurse \\
diastolic.blood.pressure & diastolic blood pressure-nurse \\
physician.sys.bp.1st.read & physician systolic pressure 1st reading
\\
physician.dia.bp.1st.read & physician diastolic pressure 1st reading
\\
physician.sys.bp.2nd.read & physician systolic pressure 2nd reading \\
physician.dia.bp.2nd.read & physician diastolic pressure 2nd reading
\\
ventricular.rate.per.minute & ventricular rate per minute \\
qrs.angle & qrs angle \\ \hline
\end{tabular}%
\label{tab:response}
\end{table}

\begin{table}[tp]
\caption{Estimation results for both the lasso and SCAD penalized estimators for the FHS data. }
\vspace{0.1in}
\label{tab:real}\centering
\begin{tabular}{ccccc}
\hline
   &LASSO&SCAD&LASSO-FR&SCAD-FR\\
\hline
Varying-coefficient Model&&&&\\
error  & 0.2052 & 0.1791  &0.2512 &0.2155\\
\# SNPs & 5     &  6      & 42    & 36\\
\hline
Linear Model&&&&\\
error &0.2509  &0.2504   &0.7309 &0.7261\\
\# SNPs & 26     &  24      & 70    & 69\\
\hline
\end{tabular}
\end{table}

\clearpage\pagebreak \newpage \thispagestyle{empty}
\begin{figure}
\centerline{\includegraphics[width=5in]{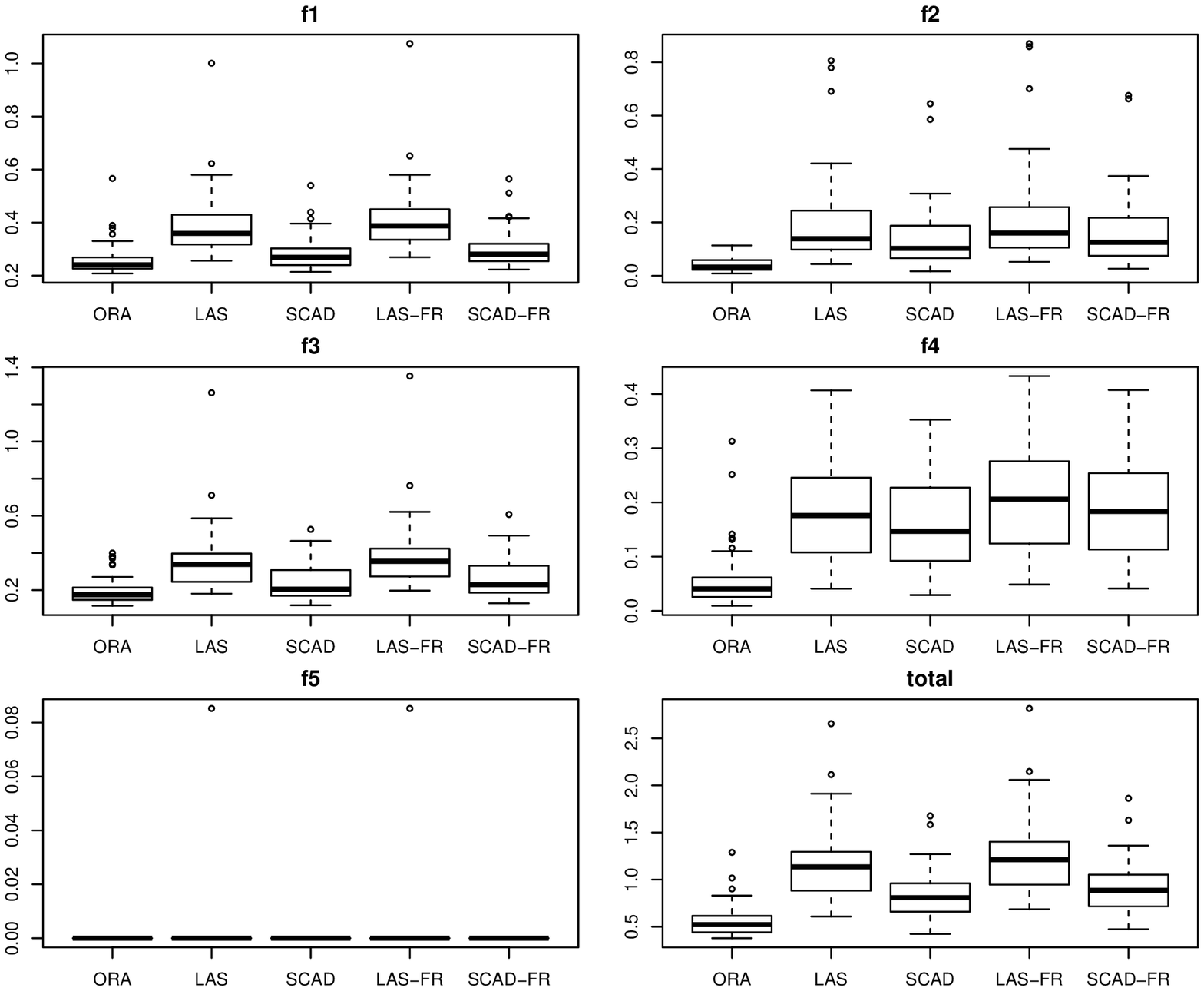}}
\caption{MSE for the estimated nonparametric functions $f_1,\ldots, f_5$ and the total MSE $\sum_{i=1}^p\|\hat{f}_j-f_j\|^2$, when $p=50$, $q=5$. \label{fig:p50q5}}
\end{figure}

\begin{figure}
\centerline{\includegraphics[width=5in]{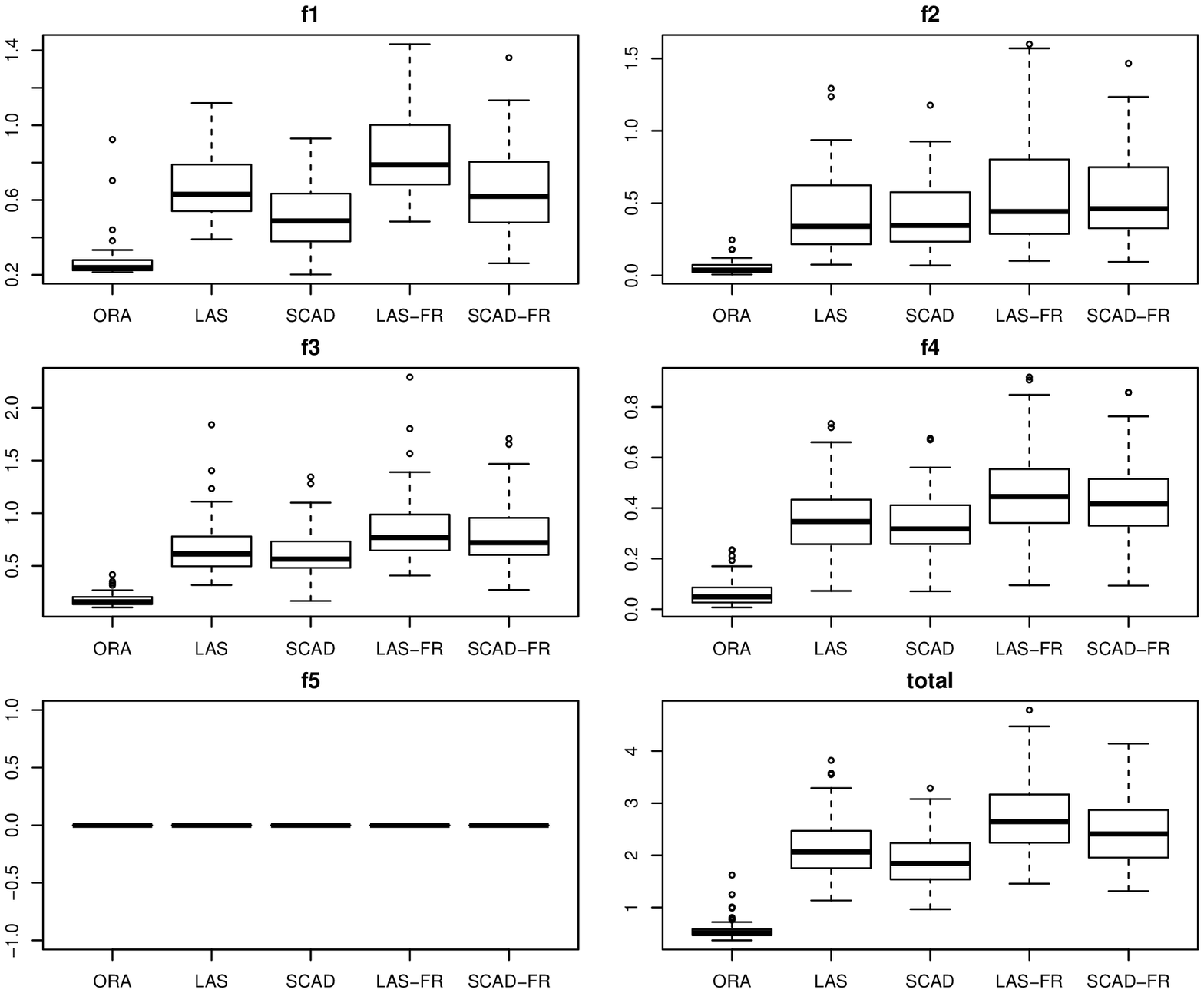}}
\caption{MSE for the estimated nonparametric functions $f_1,\ldots, f_5$ and the total MSE $\sum_{i=1}^p\|\hat{f}_j-f_j\|^2$, when $p=200$, $q=5$. \label{fig:p200q5}}
\end{figure}

\begin{figure}
\centerline{\includegraphics[width=5in]{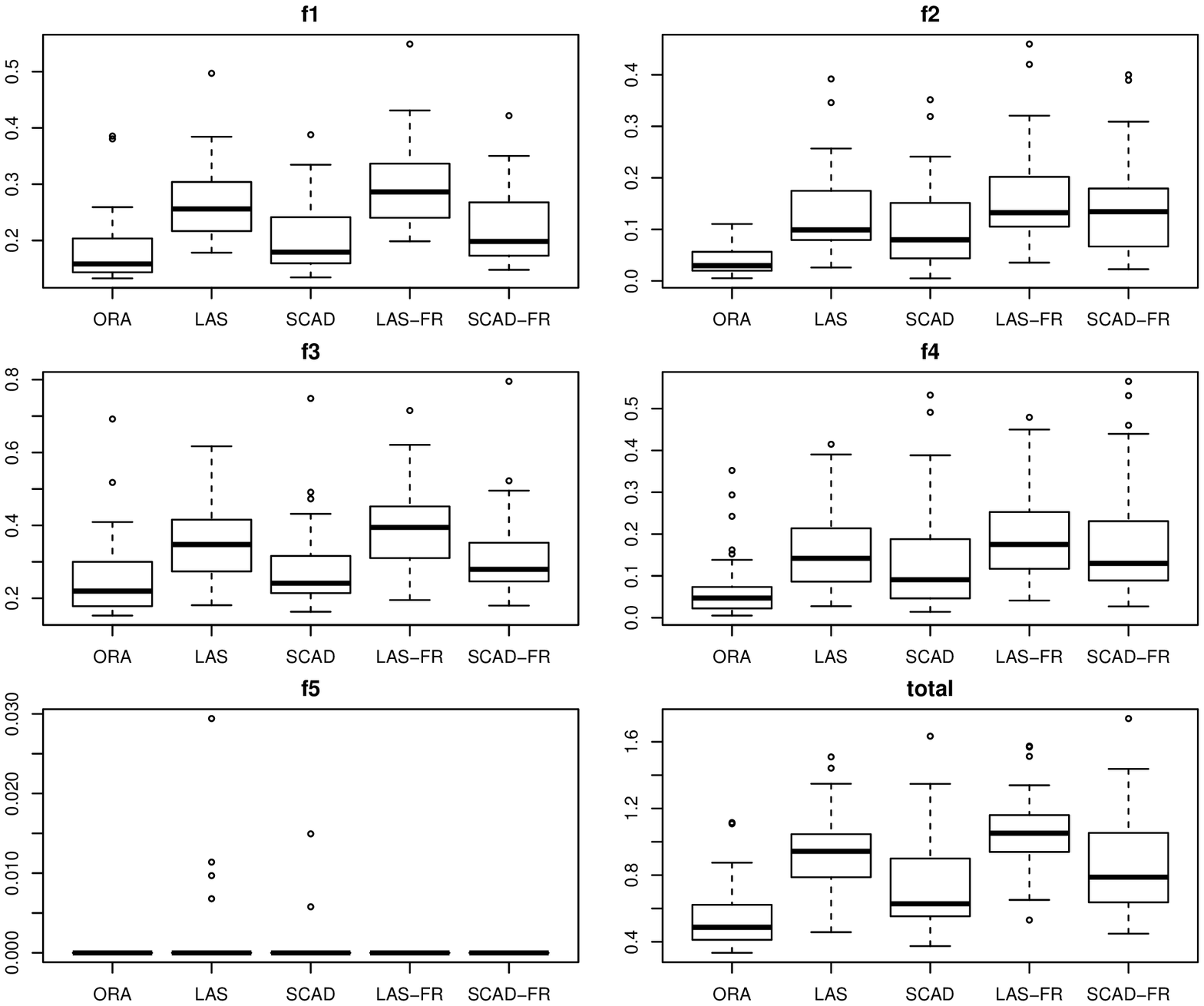}}
\caption{MSE for the estimated nonparametric functions $f_1,\ldots, f_5$ and the total MSE $\sum_{i=1}^p\|\hat{f}_j-f_j\|^2$, when $p=50$, $q=20$. \label{fig:p50q20}}
\end{figure}

\begin{figure}
\centerline{\includegraphics[width=5in]{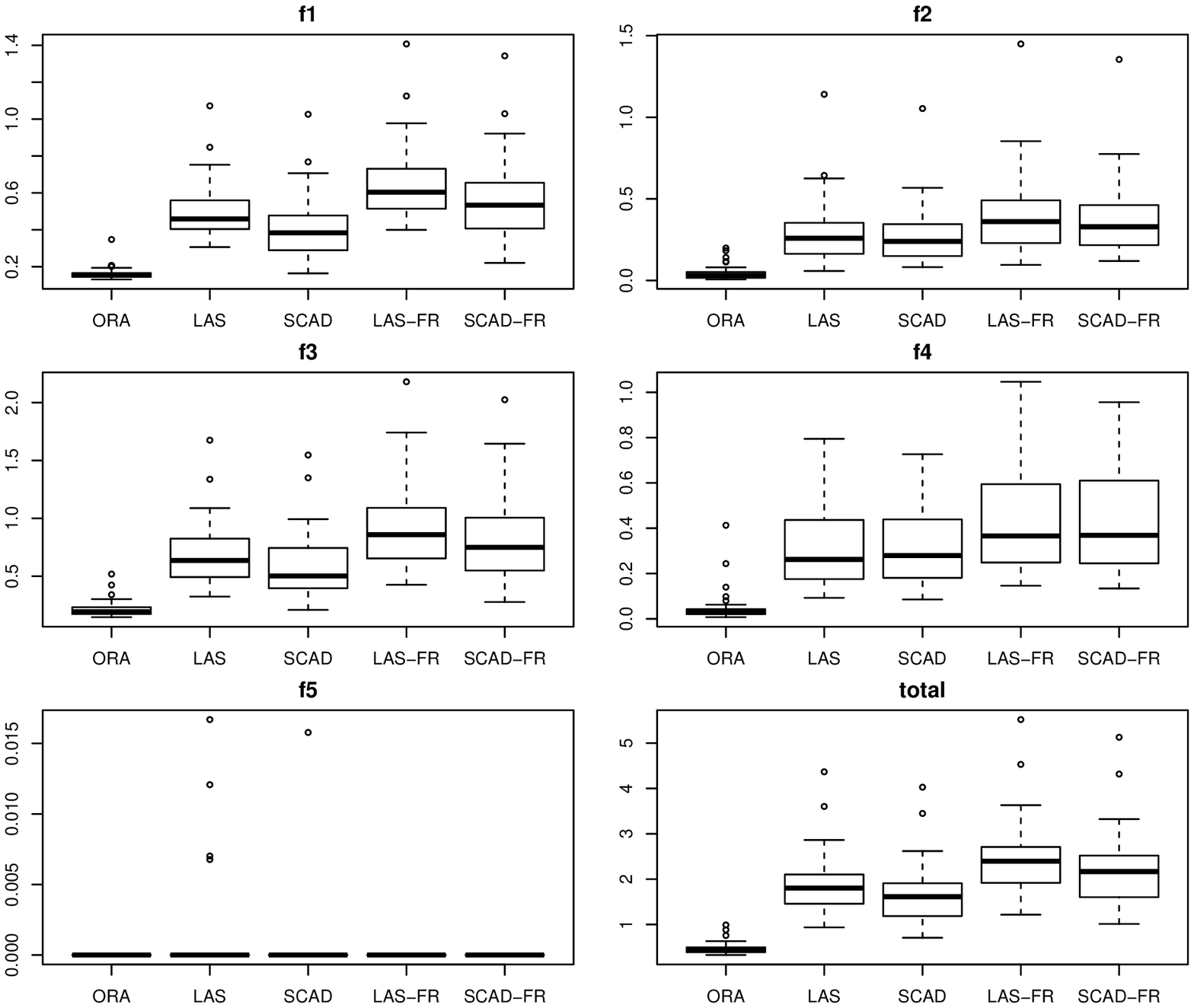}}
\caption{MSE for the estimated nonparametric functions $f_1,\ldots, f_5$ and the total MSE $\sum_{i=1}^p\|\hat{f}_j-f_j\|^2$, when $p=200$, $q=20$. \label{fig:p200q20}}
\end{figure}

\begin{figure}
\centerline{\includegraphics[width=5in]{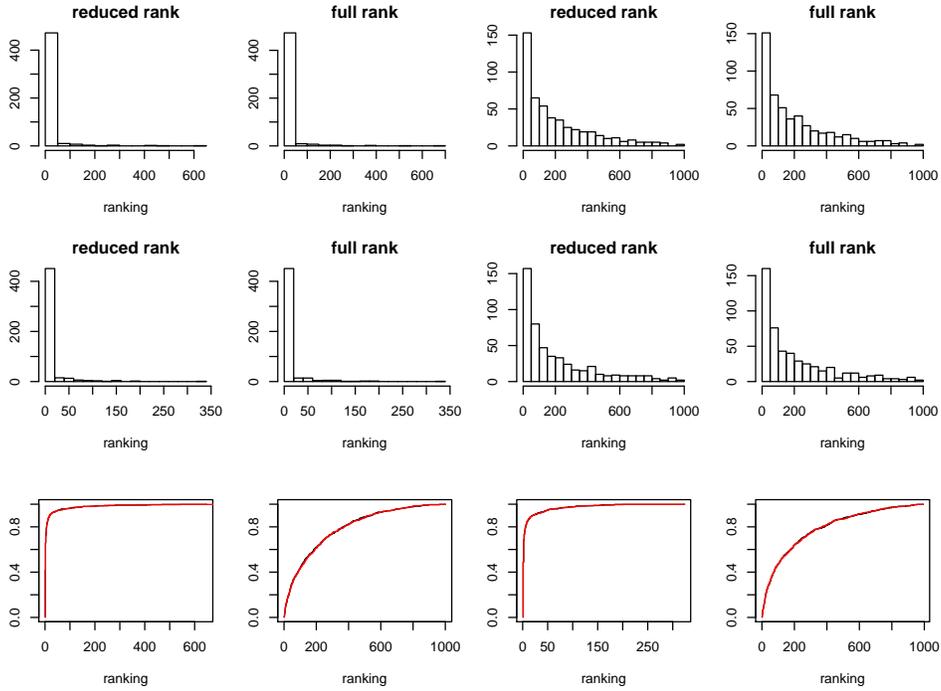}}
\caption{The first two rows show the rankings of the first four covariates resulting from screening, comparing reduced-rank screening with full-rank screening. The last row shows the empirical distribution of the rankings for  the four covariates, where the black curve is the distribution of rankings for reduced-rank screening and the red curve is the distribution of rankings for full-rank screening. The standard deviation of the Gaussian noises is $0.5$. \label{fig:screen1}}
\end{figure}

\begin{figure}
\centerline{\includegraphics[width=5in]{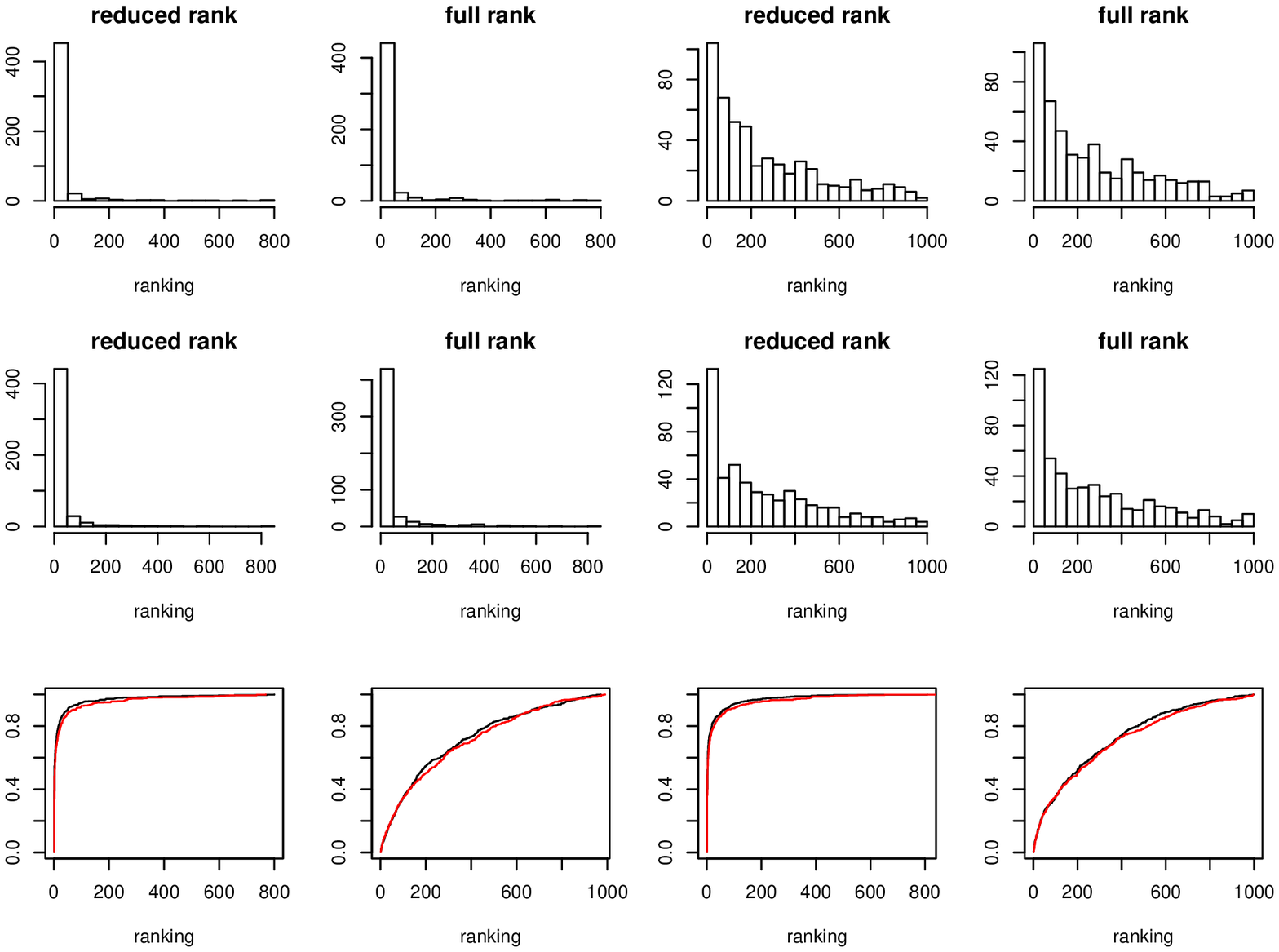}}
\caption{Similar to Figure \ref{fig:screen1} with the standard deviation of the Gaussian noises increased to $1$. \label{fig:screen2}}
\end{figure}

\begin{figure}
\centerline{\includegraphics[width=5in]{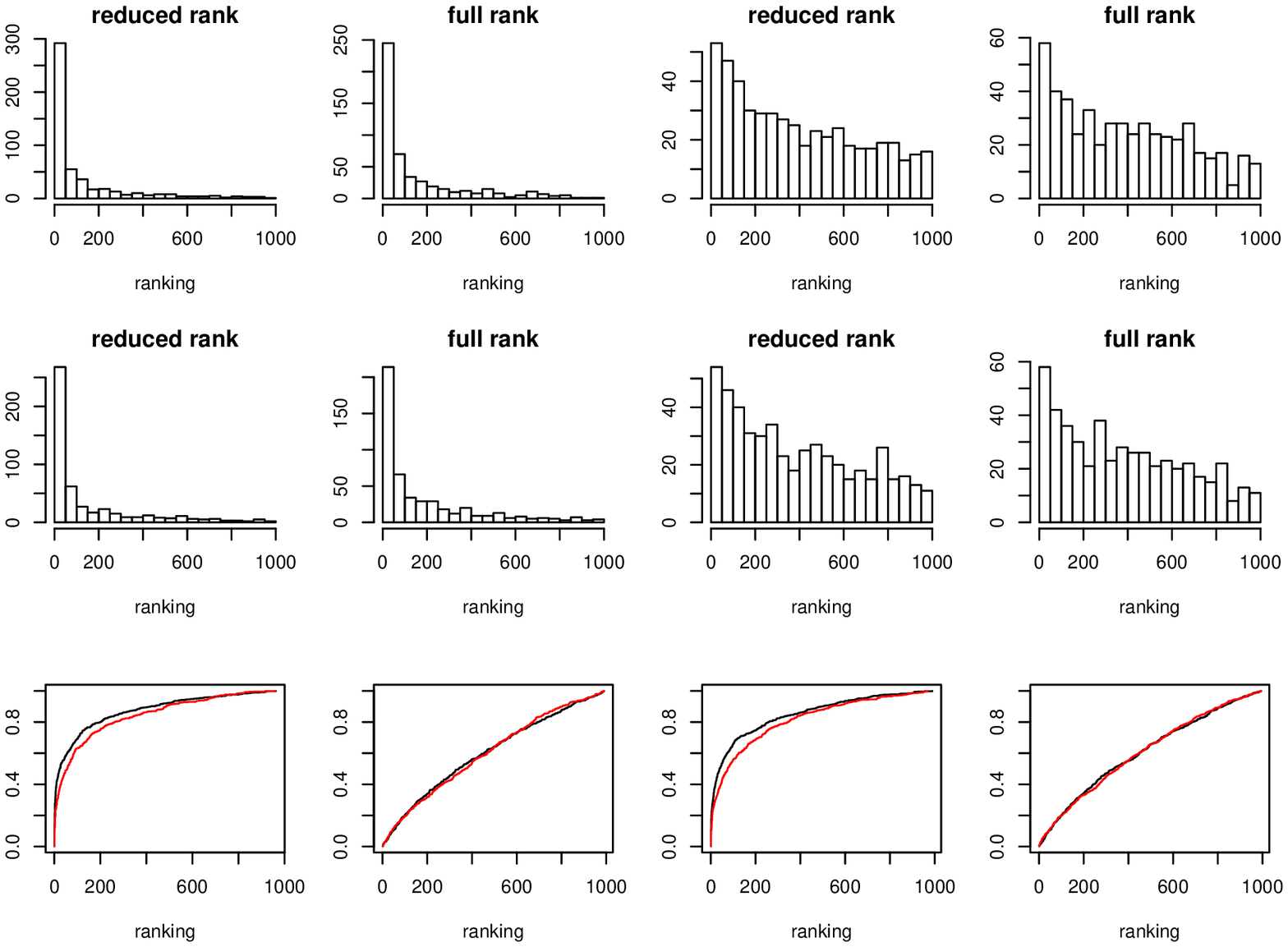}}
\caption{Similar to Figure \ref{fig:screen1} with the standard deviation of the Gaussian noises increased to $2$. \label{fig:screen3}}
\end{figure}

\begin{figure}
\centerline{\includegraphics[width=5in]{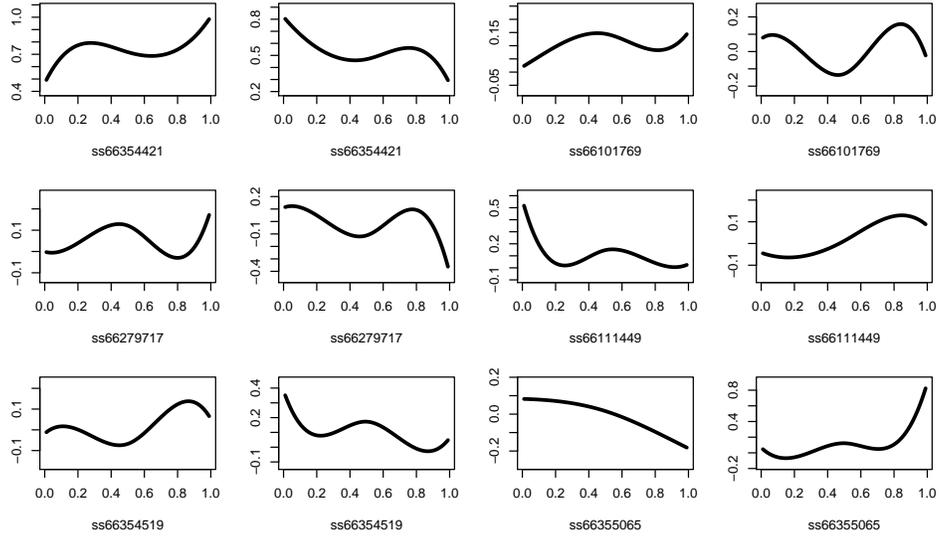}}
\caption{Estimated varying coefficients for the real data. \label{fig:real}}
\end{figure}
\end{document}